%% file: main.tex
\pdfoutput=1
\documentclass{IEEEtran}

\input{header/packages.tex}
\input{header/tensor_notations_and_misc.tex}

\begin{document}

\input{content/frontmatter.tex}

\input{content/sec_intro.tex}

\input{content/sec_MTI_PS.tex}

\input{content/sec_efficient_linearization.tex}

\input{content/sec_gen_eig}

\input{content/sec_model_case_study.tex}

\input{content/sec_results_case_study.tex}

\input{content/sec_conclusions.tex}

\input{content/backmatter.tex}

\end{document}

%% file: header/packages.tex
\usepackage[T1]{fontenc}
\usepackage[utf8]{inputenc}
\usepackage{lmodern}
\usepackage{textcomp}

\usepackage{verbatim}

\usepackage{acro} 
\input{header/acronyms.tex} 
\acsetup{locale/display,locale/format=\emph} 

\usepackage{amsmath,amssymb,amsfonts}
\usepackage{mathtools}
\usepackage{commath}
\usepackage{bm}
\usepackage[exponent-product = \cdot, output-product = \cdot]{siunitx}
\DeclareSIUnit\pu{p.u.}
\DeclareSIUnit \var {var}
\usepackage{textgreek}
\usepackage{upgreek}
\usepackage{url}
\usepackage{cite}
\usepackage{orcidlink}  
\usepackage{cleveref} 
\usepackage{graphicx}
\usepackage{stfloats}
\usepackage{subcaption}
\graphicspath{{figures/}}
\usepackage{tabularx}           
\usepackage{booktabs} 
\usepackage{multirow}

\newtheorem{definition}{Definition}[section]
\newtheorem{example}{Example}[section]

\usepackage{hyperref}
\hypersetup{
	pdfauthor={Christoph Kaufmann},
	pdftitle={Small-Signal Stability Analysis of Power Systems by Implicit Multilinear Models},
	pdfkeywords={multilinear models, power systems stability, eigenvalue analysis, converter-dominated grids}
}


%% file: header/acronyms.tex
 \DeclareAcronym{MTI}{short=MTI,long=multilinear time-invariant,short-indefinite=an}
 \DeclareAcronym{eMTI}{short=eMTI,long=explicit multilinear time-invariant,short-indefinite=an}
\DeclareAcronym{iMTI}{short=iMTI,long=implicit multilinear time-invariant,short-indefinite=an}
\DeclareAcronym{sMTI}{short=sMTI,long=semi-explicit multilinear time-invariant,short-indefinite=an}
\DeclareAcronym{ODE}{short=ODE,long=ordinary differential equation,short-indefinite=an}
\DeclareAcronym{pem}{short=PEM,long=polymer exchange membrane,short-indefinite=a}
\DeclareAcronym{ocp}{short=OCP,long=open circuit potential,short-indefinite=an}
\DeclareAcronym{LES}{short=LES,long=local energy systems}
\DeclareAcronym{CP}{short=CP,long=canonical polyadic}
\DeclareAcronym{CPN}{short=CPN,long=canonical polyadic normalized}
\DeclareAcronym{DAE}{short=DAE,long=differential-algebraic equation}
\DeclareAcronym{LTI}{short=LTI,long=linear time-invariant,short-indefinite=an}
\DeclareAcronym{CIG}{short=CIG,long=converter-interfaced generation,short-indefinite=a}
\DeclareAcronym{GFL}{short=GFL, long=grid-following,short-indefinite=a}
\DeclareAcronym{GFM}{short=GFM, long=grid-forming,short-indefinite=a}
\DeclareAcronym{SG}{short=SG, long=synchronous generator,short-indefinite=a}
\DeclareAcronym{TSO}{short=TSO, long=transmission system operator,short-indefinite=a}
\DeclareAcronym{DSO}{short=DSO, long=distribution system operator,short-indefinite=a}
\DeclareAcronym{PLL}{short=PLL, long=phase-locked loop,short-indefinite=a}
\DeclareAcronym{EMT}{short=EMT, long=electromagnetic transients,short-indefinite=an}
\DeclareAcronym{BDF}{short=BDF, long=backward differentiation formulae,short-indefinite=a}
\DeclareAcronym{POI}{short=POI, long=point of interconnection,short-indefinite=a}
\DeclareAcronym{PI}{short=PI, long=proportional-integrator,short-indefinite=a}
\DeclareAcronym{LDSS}{short=LDSS, long=linear descriptor system,short-indefinite=an}
\DeclareAcronym{CPN1}{short=CPN1,long=canonical polyadic norm-1 normalized}
\DeclareAcronym{GEP}{short=GEP,long=generalized eigenvalue problem}
\DeclareAcronym{SRF}{short=SRF,long=synchronous reference frame}
\DeclareAcronym{VSM}{short=VSM,long=virtual synchronous machine}
\DeclareAcronym{ccGFM}{short=CC-GFM,long=current control grid-forming mode}
\DeclareAcronym{VA}{short=VA,long=virtual admittance,short-indefinite=a}
\DeclareAcronym{CC}{short=CC,long=current control,short-indefinite=a}
\DeclareAcronym{SCR}{short=SCR,long=short circuit ratio,short-indefinite=an}
\DeclareAcronym{NTI}{short=NTI,long=nonlinear time-invariant,short-indefinite=an}
\DeclareAcronym{LSS}{short=LSS,long=linear state-space model,short-indefinite=an}

%% file: header/tensor_notations_and_misc.tex
\renewcommand{\vec}[1]{\mathbf{#1}} 
\newcommand{\mat}[1]{\mathbf{#1}} 
\def\tens#1{\relax\ifmmode\mathsf{#1}\else\textsf{#1}\fi} 

\newcommand{\cprod}[3]{\left\langle\,#1\,\left|\,#2\,\right.\right\rangle_{#3}}

\newcommand{\T}{\mathrm{T}}

\DeclareMathOperator{\spn}{span}

\newcommand{\MstateVar}{z}
\newcommand{\MstateVarVec}{\mathbf{z}}

\newcommand{\stateVar}{x}
\newcommand{\stateVarVec}{\mathbf{x}}

\newcommand{\newVar}{z} 
\newcommand{\newVarVec}{\mathbf{z}} 
\newcommand{\inputVar}{u} 
\newcommand{\inputVarVec}{\mathbf{u}}
\newcommand{\auxVar}{\alpha} 
\newcommand{\auxVarVec}{\bm{\alpha}}

\newcommand{\paramMatrix}{\mathbf{\Phi}} 
\newcommand{\structureMatrix}{\mathbf{S}} 

\newcommand{\variableVector}{\mathbf{v}} 

\newcommand{\structuredCP}{\mathbf{s}}

\newcommand{\TrfVars}{\bm{\sigma}} 

%% file: content/frontmatter.tex
\title{Small-Signal Stability Analysis of Power Systems by Implicit Multilinear Models} 
\author{Christoph Kaufmann\,\orcidlink{0000-0002-0666-1104}, Georg Pangalos\,\orcidlink{0000-0001-5094-8033}, Gerwald Lichtenberg\,\orcidlink{0000-0001-6032-0733}, Oriol Gomis-Bellmunt\,\orcidlink{0000-0002-9507-8278}~\IEEEmembership{Fellow,~IEEE}
\thanks{Manuscript received August XX, XXXX; revised August XX, XX.(\textit{Corresponding author: Christoph Kaufmann})}%
\thanks{This work was supported in part by the European Union Horizon Europe program under Grant Agreement no. 101096197 (project AGISTIN), and in part funded by CETPartnership, the Clean Energy Transition Partnership under the 2023 joint call for research proposals, co-funded by the European Commission (GA 101 069750) and with the funding organizations detailed on~\url{https://cetpartnership.eu/funding-agencies-and-callmodules} including the BMWE under Grant Agreement no. 03EI4095 in the TenSyGrid project, and in part by the ICREA Academia program. 
}%
\thanks{Christoph Kaufmann is with the Application Center for Integration of Local Energy Systems ILES, Fraunhofer Institute for Wind Energy Systems IWES, 27572 Bremerhaven, Germany, also with the Faculty Life Sciences, Hamburg University of Applied Sciences, 21033 Hamburg, Germany, and also with Centre d’Innovació Tecnològica en Convertidors Estàtics i Accionaments, Departament d’Enginyeria Elèctrica, Universitat Politècnica de Catalunya, 08028 Barcelona, Spain (e-mail: christoph.kaufmann@iwes.fraunhofer.de).}%
\thanks{
Georg Pangalos is with the Application Center for Integration of Local Energy Systems ILES, Fraunhofer Institute for Wind Energy Systems IWES, 27572 Bremerhaven, Germany, also with the Faculty Life Sciences, Hamburg University of Applied Sciences, 21033 Hamburg, Germany (e-mail: georg.pangalos@iwes.fraunhofer.de). 
}%
\thanks{Gerwald Lichtenberg is with the Faculty Life Sciences, Hamburg University of Applied Sciences, 21033 Hamburg, Germany (e-mail: gerwald.lichtenberg@haw-hamburg.de).}
\thanks{
Oriol Gomis-Bellmunt is with the Centre d’Innovació Tecnològica en Convertidors Est\`atics i Accionaments,
Departament d’Enginyeria Elèctrica, Universitat Polit\`ecnica de Catalunya, 08028 Barcelona, Spain (e-mail: oriol.gomis@upc.edu).
}%
}
\markboth{Journal of \LaTeX\ Class Files,~Vol.~14, No.~8, August~2021}%
{Shell \MakeLowercase{\textit{et al.}}: A Sample Article Using IEEEtran.cls for IEEE Journals}
\maketitle
\begin{abstract}
This paper proposes a new approach to perform small-signal stability analysis based on linearization of implicit multilinear models. 
Multilinear models describe the system dynamics by multilinear functions of state, input, and algebraic variables. Using suitable transformations of variables, they can also represent trigonometric functions, which often occur in power systems modeling. 
This allows tensor representations of grid-following and grid-forming power converters. 
This paper introduces small-signal stability analysis of equilibrium points based on implicit multilinear models using generalized eigenvalues.
The generalized eigenvalues are computed from linear descriptor models of the linearized implicit multilinear model. 
The proposed approach is tested using a 3-bus network example, first by comparing time-domain simulations of the implicit multilinear model with those of the nonlinear model, and second by comparing the generalized eigenvalues with those of the linearized nonlinear model. 
The results show that the decomposed tensor representation of the implicit multilinear model allows for a faster linearization compared to conventional methods in MATLAB Simulink. 
\end{abstract}


%% file: content/sec_intro.tex
\section{Introduction}\label{sec:intro}
\IEEEPARstart{P}{ower} system stability analysis is challenging because power systems are hybrid nonlinear dynamical systems~\cite{Kundur.2004,Fourlas.2004}, i.e., a mixture of nonlinear continuous and discrete-valued dynamics has to be coped with.
There are no closed-form solutions to all types of hybrid nonlinear differential equations~\cite{Khalil.2015} that describe the behavior of power systems.  

Approximative solutions can be obtained by time-domain simulations using numerical integrations, which is the most common approach~\cite{Kundur.2004,F.Milano.2018}. 
 However, the computational burden can be high, especially for converter-dominated, low-inertia networks~\cite{Farrokhabadi.2020,Markovic.2021}, and they only allow for a case-by-case analysis, requiring numerous simulations, where general conclusions are difficult to be drawn~\cite{F.Milano.2018}. 

Analytic method can directly assess the stability of a system~\cite{Khalil.2015}. 
They can provide useful insights and conclusions about the system's behavior, but they are restricted to specific structures of the governing differential and algebraic equations,~i.e., the model class~\cite{F.Milano.2018,Kundur.2004,Khalil.2015}.
Lyapunov's direct method allows for studying large-signal power system stability. However, there is no general framework to construct Lyapunov functions for nonlinear systems. Assumptions are needed to simplify the model, which limit their applications to real-world power systems~\cite{Kundur.2004,Farrokhabadi.2020}. 
An example where the restriction to a model class helped is the work~\cite{Anghel.2013}, where an algorithmic construction of Lyapunov functions of a multi-machine system could be formulated by finding of polynomial model representation of the synchronous generator. 

The most-used analytic method in power system stability analysis is approximating the nonlinear model as a linear model locally at an equilibrium point, and assessing the small-signal stability~\cite{Kundur.2004,Hatziargyriou.2020,Farrokhabadi.2020}.
The small-signal stability analysis of converter-dominated networks can be either done in time-domain by the eigenvalues of linear state-space models~\cite{Khalil.2015,Pogaku.2007,Markovic.2021}, or in frequency domain~\cite{Orellana.2021}, which allows estimating the input-output behavior of proprietary models and equipment through external excitation~\cite{Orellana.2021}. 

 This work shows how the modeling of power systems can be done by only using multilinear models~\cite{Pangalos.2013,Lichtenberg.2022}. 
Power systems encounter some per se multilinear functions, for example considering the calculation of active power,~$P$ in a three-phase system in~$abc$-coordinates by
    \begin{IEEEeqnarray}{rCL}
         \label{eq:PQ_calc}
        p(t)&=&v_a(t)i_a(t)+v_b(t)i_b(t)+v_c(t)i_c(t)\,, \label{eq:P_calc}
    \end{IEEEeqnarray}
    where the multiplication of the voltage~$v$ and the current~$i$ is a multilinear function,~i.e., it is linear if all but one variable are held constant~\cite{Pangalos.2015}.

    Multilinear models have been an active field of research for more than the last decade and applied to heating systems~\cite{Pangalos.2013,LeonaSchnelle.2022}, multi-agent systems~\cite{Pangalos.2013}, 
     chemical systems~\cite{Kruppa.2014}, and more recently to electrolyzers~\cite{Lichtenberg.2022}, and power systems~\cite{Kaufmann.2023,Samaniego.2024}. 
     Multilinear models are a subclass of polynomial models, that can also represent boolean dynamics besides linear ones~\cite{Pangalos.2013,Lichtenberg.2022}. 
  The representation of multilinear models as decomposed tensors allows efficient representation of the model~\cite{Pangalos.2013,Jores.2022}, as well as efficient linearization~\cite{Kaufmann.2023b}, motivating the use for high dimensional and complex power system models. 

 Modeling power systems as multilinear models was firstly done in~\cite{Kaufmann.2023}, where a $PQ$ open-loop controlled photovoltaic inverter in~$\alpha\beta$-coordinates was modeled.
 Power systems encounter nonlinear functions, like trigonometric functions for example. These were firstly represented as multilinear model in~\cite{Samaniego.2024}, which modeled a larger converter-based microgrid in $dq$-frame, where each converter used a droop-based control. 
 This work builds upon the ongoing research and extends it by the following contributions:  
\begin{itemize}
    \item Extending the efficient linearization procedure for~\acl{eMTI} models presented in~\cite{Kaufmann.2023b} to~\acl{iMTI} models leading to local approximations by linear descriptor state-space models.
     \item Applying the small-signal stability analysis method for models described by nonlinear~\aclp{DAE} using generalized eigenvalues~\cite{Milano.2020} to the model class of~\acl{iMTI} models.
     \item Exact symbolic multilinearization of relevant non-multilinear functions for the modeling of grid-following and grid-forming power converters.
  \end{itemize}
  It will be shown that the proposed methodology allows for an efficient small-signal stability analysis of power systems compared to standard tools. 

The paper is organized as follows. 
Section~\ref{sec:MTI} describes the class of~\acl{iMTI} models and its representation as decomposed tensors. 
Exact multilinear modeling of some nonlinear functions encountered in power converters are discussed in Section~\ref{sec:symPC}. 
Then Section~\ref{sec:effLin} shows the efficient computation of the Jacobian of~\ac{iMTI} models, out of which a linear descriptor model is extracted.
From the linear descriptor model, the generalized eigenvalues are computed as presented in Section~\label{sec:GEP}.
Section~\ref{sec:ps_ex} outlines the model used in Section~\ref{sec:results}, for the comparison of the time-domain simulation and small-signal analysis using Matlab Simulink~\cite{MATLAB.2023}. 
Section~\ref{sec:con} draws conclusion of the work and defines the next steps.

%% file: content/sec_MTI_PS.tex
\section{Implicit Multilinear Models and Tensor Representation} \label{sec:MTI}
\Ac{MTI} models were introduced as in~\cite{Pangalos.2013} as state space models with multilinear functions as right-hand sides. 
These~\ac{MTI} state-space models are also referred to as~\ac{eMTI} models because their dynamic behavior is described by~\mbox{\acp{ODE}}. 
If the~\acp{ODE} are restricted by a set of algebraic equations, meaning the model is described by~\acp{DAE}, the~\ac{sMTI}~model class~\cite{Kaufmann.2023} can be used. Writing the~\mbox{\acp{DAE}} fully implicitly leads to the~\ac{iMTI} model class~\cite{Lichtenberg.2022}. 

In this paper, the set of real numbers is denoted by~$\mathbb{R}$, where a scalar parameter is indicated by a non-bold face~$x\in\mathbb{R}$, lower case bold face is used to indicate a column vector~$\stateVarVec\in\mathbb{R}^n$, upper case bold face for matrices with~$\mathbf{M}\in \mathbb{R}^{n\times m}$, and upper case non-bold face for a $k$-th order tensor~$\tens{T}\in\mathbb{R}^{I_1 \times I_2\times \hdots \times I_k}$. 

In the subsequent section, the representation of~\ac{iMTI} models as decomposed tensors is briefly presented starting with the full tensor representation.   

To formulate an~\ac{iMTI} state-space model, the monomial tensor is defined at first.
\begin{definition}~\cite{Kolda.2009,Pangalos.2013,Lichtenberg.2022} A~\textit{monomial tensor} is defined as 
\begin{IEEEeqnarray*}{rCl} 
    &&\tens{M}\left(\dot{\MstateVarVec}, \MstateVarVec, \vec{u},\vec{y},\auxVarVec\right)\\
    &&=\!\begin{pmatrix} 1 \\\dot{\MstateVar}_1\end{pmatrix}\! \circ\!\hdots\!\circ\!\begin{pmatrix} 1 \\ \dot{\MstateVar}_n\end{pmatrix}\!\circ\! \begin{pmatrix} 1 \\ {\MstateVar}_1\end{pmatrix}\circ\!\hdots \!\circ\!\begin{pmatrix} 1 \\ {\MstateVar}_n\end{pmatrix}\! \circ\! \begin{pmatrix} 1 \\ {u}_1\end{pmatrix}\!\circ\!\hdots\!\circ\! \begin{pmatrix} 1 \\ {u}_m\end{pmatrix}\\
    &&\hdots\!\circ\! \begin{pmatrix} 1 \\ {y}_1\end{pmatrix}\!\circ\!\hdots\!\circ\! \begin{pmatrix} 1 \\ {y}_p\end{pmatrix}\!\circ\!  \begin{pmatrix} 1 \\ {\auxVar}_1\end{pmatrix}\! \circ\! \hdots\!\circ\! \begin{pmatrix} 1 \\ {\auxVar}_q\end{pmatrix}\in\mathbb{R}^{\overbrace{ ^{2 \times \ldots \times 2} }^{2n+m+p+q}}\,,
\end{IEEEeqnarray*}
where the outer product provides all possible multilinear combinations of the~$n\in\mathbb{N}$~state variables~\mbox{$\MstateVarVec\in\mathbb{R}^n$},~\mbox{$m\in\mathbb{N}$}~input variables~\mbox{$\vec{u}\in\mathbb{R}^m$},~\mbox{$p\in\mathbb{N}$}~output variables~\mbox{$y\in\mathbb{R}^p$}, and~\mbox{$q\in\mathbb{N}$} algebraic variables~\mbox{$\auxVarVec\in\mathbb{R}^{q}$}. The monomial tensor is structurally of rank 1,~i.e., there is only one outer product and not a sum of two or more necessary for representation~\cite{Kolda.2009}. 
\end{definition}

\begin{example}\label{ex:monomial}
    Consider an~\ac{iMTI} model with $n=1$ state, and~$q=1$ algebraic variable. Thus, there are~$2^3=8$  combinations for multilinear monomials, and the resulting monomial tensor is depicted in Fig.~\ref{fig:monomial}.  
    
    \begin{figure}[h]
        \centering
        \includegraphics{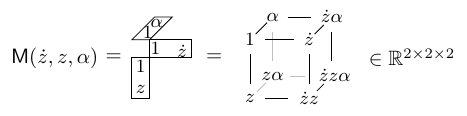}
        \caption{Monomial tensor of the Example~\ref{ex:monomial}}
        \label{fig:monomial}
    \end{figure}
\end{example}
The total number of relevant signals of the model is~\mbox{$N_v=2n+m+p+q$}.  
In the following the notation for tensor spaces is used~\cite{Lichtenberg.2022}
 \begin{IEEEeqnarray*}{rCl}
    \mathbb{R}^{\times^{N_v}2}:=\mathbb{R}^{\overbrace{ ^{2 \times \ldots \times 2} }^{N_v}}\,. 
\end{IEEEeqnarray*}
The multilinear model equations are obtained from their tensors using the contracted product. 
\begin{definition}~\cite{Lichtenberg.2022,Kolda.2009}
    The~\textit{contracted product}~$\cprod{\cdot}{\cdot}{}$ of a tensor~\mbox{$\tens{F}\in\mathbb{R}^{I_1 \times \cdots \times I_k \times J_1 \cdots  J_l}$}, and a tensor~\mbox{$\tens{G} \in \mathbb{R}^{I_1 \times \cdots \times I_k}$}, gives a tensor~\mbox{$\tens{H}=\cprod{\tens{F}}{\tens{G}}{}~\in\mathbb{R}^{J_1\times \cdots \times J_l}$} with elements
\begin{IEEEeqnarray}{rCl}
     h_{j_1,\hdots,j_l}&& =\sum_{i_1=1}^{I_1}\sum_{i_2=1}^{I_2}\!\!\cdots\!\! \sum_{i_k=1}^{I_k}g_{i_1, \hdots, i_k}f_{i_1, \hdots, i_k,j_1, \hdots, j_l}\,, 
\end{IEEEeqnarray}
where the elements are indexed by~\mbox{$j_l\in\{1,2,\hdots,J_l\}$ for $l=1,\hdots,J_l$}. 
\end{definition}
With the monomial tensor and the contracted product, an~\ac{iMTI} model can be defined.
\begin{definition}~\cite{Lichtenberg.2022} An~\acl{iMTI} state-space model is defined as
\begin{IEEEeqnarray}{rCl}
   \vec{0}=\cprod{\tens{H}}{\tens{M}\left(\dot{\MstateVarVec}, \MstateVarVec, \vec{u},\vec{y}, \auxVarVec\right)}{} \in\mathbb{R}^{n+Q+p} \,, \label{eq:iMTIcp}
\end{IEEEeqnarray}
where the parameter tensor has the dimensions~\mbox{$\tens{H}\in\mathbb{R}^{\times^{N_v}2\times {(n+Q+p)}}$}, since there are $n$-state equations of the model that are constrained by~\mbox{$Q\in\mathbb{N}$}~algebraic equations, and it has~$p$ output equations. The total number of equations is~\mbox{$N_\mathrm{\phi}=n+Q+p$}. The right-hand side of the~\ac{iMTI} model is described by the implicit multilinear function 
\begin{IEEEeqnarray}{rCl}
    \vec{0} &=&\vec{h}(\dot{\MstateVarVec},\MstateVarVec,\vec{u},\vec{y},\auxVarVec)\,,
    \label{eq:impFcn}   
\end{IEEEeqnarray}
where $\vec{h}: \mathbb{R}^n \times \mathbb{R}^n \times \mathbb{R}^m \times \mathbb{R}^p \times \mathbb{R}^q \to  \mathbb{R}^{N_\mathrm{\phi}}$. 
\end{definition}

The full parameter tensor~$\tens{H}$ of an~\ac{iMTI} model suffers from the curse of dimensionality since the number of elements are~$N_\mathrm{\phi}2^{N_v}$. 
To overcome the curse of dimensionality, the tensor is decomposed using a normalized format of~\ac{CP} tensor decomposition~\cite{Kolda.2009}, called~\ac{CPN1} format~\cite{LeonaSchnelle.2022}. 
The idea of the~\ac{CP} decomposition is to represent a tensor~$\tens{F}$ by a finite sum of rank-one tensors, where the rank-one tensors are obtained by the outer product as shown in Example~\ref{ex:monomial}~\cite{Hitchcock.1927,Kolda.2009}. 
In~\cite{Hitchcock.1927}, the~\textit{rank} of the tensor~$\tens{F}$ is defined as the smallest number of rank-one tensors required to obtain an exact representation of~$\tens{F}$~\cite{Kolda.2009}. 
Determining the rank of a tensor~$\tens{F}$ is an NP-hard problem~\cite{Hastad.1990,Kolda.2009}. 
However, there are algorithms that can find low-rank approximations that fulfill certain accuracy requirements~\cite{Kolda.2009}. 
This work only considers exact representations with presumably non-minimal rank. 
In the context of multilinear models~\cite{Pangalos.2013}, this means the representation is not necessarily minimal. 
The~\ac{CPN1} format allows for another reduction of the number of elements to the factors by almost half~\cite{LeonaSchnelle.2022,Jores.2022}. 
Using the~\ac{CPN1} format, the number of elements of the parameter tensor scale linearly with number of factors~$R \in\mathbb{N}$ and the number of dimensions, i.e., it has~\mbox{$N_\mathrm{\phi} R$} elements, in contrast to the exponential increase~$N_\mathrm{\phi}2^{N_v}$ of the full parameter tensor~$\tens{H}$ of the~\ac{iMTI} model.

\begin{definition}~\cite{Kaufmann.2023, Kaufmann.2023b}
The contracted product of an~\ac{iMTI} model in the \ac{CPN1} format is written as a simple sum of factored multilinear polynomials
\begin{IEEEeqnarray}{rCl}
    \IEEEyesnumber \label{eq:iMTI_CPN1}
   \vec{0}&=&\paramMatrix\,\vec{s}(\vec{v}(t))\,,\IEEEyessubnumber\\
  \vec{s}_r(\vec{v}(t)) &=&\prod_{i=1}^{N_v}{\left(1\!-|\structureMatrix_{ir}|\,\!+\!\structureMatrix_{ir}\variableVector_i(t)\right)}\,,\IEEEyessubnumber \label{eq:structuredCP}
\end{IEEEeqnarray}
where~$\mathbf{v}(t)=\left(\dot{\MstateVarVec}(t)^\T, \MstateVarVec(t)^\T, \vec{u}(t)^\T,\vec{y}(t)^\T,\auxVarVec(t)^\T\right)^\T$ is a column vector containing all states, inputs, outputs, and algebraic variables. 
The dependence on time is explicitly stated here to highlight that the structured contracted product at time~$t$~is a vector-valued function $\structuredCP:\mathbb{R}^{N_v}\to\mathbb{R}^{R}$, where~$R\in\mathbb{N}$ is the number of factors of the parameter tensor of the~\ac{iMTI} model.
In contrast, the time-invariant parameter matrix~\mbox{$\paramMatrix\in\mathbb{R}^{N_\mathrm{\phi}\times R}$} gives the coefficients of the~\ac{iMTI} model. 
The structural matrix~\mbox{$\structureMatrix\in\mathbb{R}^{N_v\times R}$} provides the factors for the multilinear combinations of the variables in~$\vec{v}$. 
\end{definition}

To illustrate the usage of the~\ac{CPN1} format for multilinear models, one example is provided. 

\begin{example}
 The calculation of the active power~$p$ is a multilinear function~\eqref{eq:P_calc}, and can be written implicitly as
 \begin{IEEEeqnarray}{rCl}
 0=p-v_a i_a-v_b i_b-v_c i_c\,.\label{eq:P_implicit}
 \end{IEEEeqnarray}
 It has six inputs~\mbox{$\mathbf{u}=\begin{pmatrix}
    v_a, & v_b, &v_c, &i_a, & i_b, & i_c 
 \end{pmatrix}^\T\in\mathbb{R}^{6}$}, and one output~\mbox{$y=p\in\mathbb{R}$}. The parameter matrix~$\paramMatrix$ provides the coefficients of the 4 terms of~\eqref{eq:P_implicit}, in this case leading to~$R=4$, thus
    \begin{IEEEeqnarray*}{rCl}
       \paramMatrix&=&\begin{pmatrix}
            1,&-1, & -1,&-1
            \end{pmatrix}\,,
        \end{IEEEeqnarray*} 
where each column corresponds to the combination of the variables provided by the structural matrix~$\structureMatrix$. In the structural matrix, the rows correspond to each variable, in this case the elements of~$\left(\mathbf{u}^\T,y\right)^\T$ in this order, and provide the multilinear combinations of the variables used for the~\mbox{$R$-terms}, such that~$\structureMatrix\in\mathbb{R}^{7\times4}$.
The second term~$-V_a I_a$ of the implicit expression of~\eqref{eq:P_implicit} for instance, is indicated by the two nonzero entries $\structureMatrix_{12}=\structureMatrix_{42}=1$, corresponding to $\paramMatrix_{12}=-1$.
 Following this procedure, the structural matrix of~\eqref{eq:P_implicit} is
        \begin{IEEEeqnarray*}{rCl}
        \structureMatrix&=&
        \begin{pmatrix}    
                       0&1&0&0\\ 
                       0&0&1&0\\ 
                       0&0&0&1\\ 
                       0&1&0&0\\ 
                       0&0&1&0\\ 
                       0&0&0&1\\ 
                       1&0&0&0\\ 
        \end{pmatrix}\,,\quad \variableVector(t)=\begin{pmatrix}
            v_a\\ v_b \\ v_c\\ i_a \\ i_b \\ i_c \\ p
        \end{pmatrix}\,.
    \end{IEEEeqnarray*} 
    Note that this is an introductory example to familiarize the reader with the notation, where~$\structureMatrix$ consists of only~$1$ and~$0$, though it could be any real number between~$0$ and~$1$. 
    The advantage of the factorized~\ac{CPN1} format becomes more evident, when considering larger examples, where good factorizations are available, see~\cite{Jores.2022} or~\cite{Kaufmann.2023b} for more examples.  
\end{example}

The composition of~\ac{iMTI} models is straightforward~\cite{Lichtenberg.2022} and demonstrated by an example. 
\begin{example}
   Connecting two models in series is done by simply adding an algebraic constraint, for example,~\mbox{$0=y_1-u_2$}, where the output of the first model~$y_1$ is the input of the second model~$u_2$.  
\end{example}

The developed approaches presented in this paper use the MTI-Toolbox~\cite{Lichtenberg.2024} that is available for \textsc{Matlab}~\cite{MATLAB.2023}. Several examples are given in the MTI-Toolbox that demonstrate its usage, and the work~\cite{Samaniego.2024} provides an example of a nine bus power system. 

\section{Symbolic Multilinearization of Converter-relevant Non-Multilinear Dynamics}\label{sec:symPC}
There are different possibilities to obtain a multilinear model from a nonlinear power system model. 
Numerical multilinearization~\cite{Kruppa.2014} generates an approximative multilinear model by, e.g., sparse gridding of the nonlinear model. 
This paper concerns symbolic multilinearization, by which an exact multilinear representation of the nonlinear power system model is found with the help of auxiliary variables. 
Some of them are already multilinear, as demonstrated in the Section~\ref{sec:intro}. 

\subsection{Non-Multilinearities of Power Converters}
The controllers of power converters have several common nonlinear functions. For example, there are saturations in PI controllers, Euclidean norms for voltage or current magnitude calculations using $dq$-components in voltage controllers or for current saturation algorithms, trigonometric functions for Park transformations or using rotational matrices, or switching between modes of operations during,~e.g., faults,~\cite{TenSyGrid.D31.2025}. 

The nonlinear behavior of power electronics are disregarded and  ideal averaged-switching models assumed as the focus is on eigenvalue analysis. 
Other components in power networks are often approximated as linear models using lumped parameters, see for example~\cite{Orellana.2021}, such as a transformer as an $RL$-section, or the cables as $\pi$-models. 

Further examples for implicit formulations of power system components can be found in~\cite{Milano.2016}, where a synchronous machine model and several controllers are written in an semi-implicit formulation, which can be easily transferred into an~\ac{iMTI} model. 

Next, different examples are given for the symbolic multilinearization. 

\subsection{Higher-order Polynomial Models}
Polynomial models can be represented by~\ac{iMTI} models by introducing additional algebraic variables~\cite{Lichtenberg.2022}. The procedure is demonstrated by an example. 
\begin{example}
    Consider a first-order autonomous polynomial model 
    \begin{IEEEeqnarray*}{rCl}
        \dot{x}=x^2\,.
    \end{IEEEeqnarray*}
    Introducing an auxiliary algebraic variable~$\alpha$, the polynomial model can be written as an~\ac{iMTI} model
    \begin{IEEEeqnarray*}{rCl}
        0&=&\dot{x}-x\auxVar\,,\\
        0&=&x-\auxVar\,.
    \end{IEEEeqnarray*}
\end{example}

\subsection{Trigonometric Functions} \label{subsubsec:TrigFunc}
The work~\cite{Savageau.1987} showed canonical expressions of different nonlinear functions by a change of variables into a polynomial formulation.  This change of variables was used for modeling synchronous generators in~\cite{Anghel.2013,Tacchi.2018} to obtain a sum-of-squares formulation of the model. 
The representation of the nonlinear model into a polynomial model is achieved by a nonlinear change of coordinates defined by smooth mapping 
\begin{IEEEeqnarray}{rCl} \IEEEyesnumber\label{eq:Trans}\IEEEyessubnumber
 z_1&=&\cos{(x)}\,,\\
 z_2&=&\sin{(x)}\,, \IEEEyessubnumber
\end{IEEEeqnarray}
with the algebraic constraint
\begin{IEEEeqnarray}{rCl}
g(\vec{z})&=&z_1^2+z_2^2-1=0\,.
\label{eq:trigConstraint} \label{eq:PLL_dq_constraints}
\end{IEEEeqnarray}

Despite the change of the state-space dimensions, the work~\cite{Tacchi.2018} showed, that for their chosen invertible transformation, their example is a Lie-Bläcklund isomorphism~\cite{Fliess.1999}, which means that the two models have the same trajectory, and they can be called~\textit{equivalent}. The definition of a Lie-Bläcklund transformation is provided in the Appendix~\ref{sec:LBtrf}. 

The process of the change of coordinates is demonstrated using~a \ac{PLL} as an example. 
\begin{example}[PLL]\label{ex:PLL1}
To simplify the explanation, the network is assumed to be in $dq$-coordinates. 
The {PLL} is depicted in~Fig.~\ref{fig:PLL}, and the Park transformation is replaced by the rotational matrix~$\mathbf{R}(\delta): \mathbb{R}^2\to \mathbb{R}^2$
\begin{IEEEeqnarray}{rCl}
\mathbf{R}(\delta)&=&\begin{pmatrix}
\cos{(\delta)} & - \sin{(\delta)}\\
\sin{(\delta)} & \cos{(\delta)}
\end{pmatrix}\,,
\label{eq:rot_mat}
\end{IEEEeqnarray}
where~$\delta\in\mathbb{R}$ is the angle difference between the local converter synchronous frame, denoted by~$dq$, and the global rotating synchronous frame of the network denoted by~$DQ$ as in the work~\cite{Pogaku.2007}.  

    \begin{figure}
        \centering
                \includegraphics{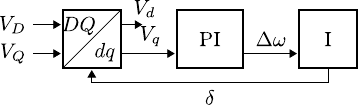}
        \caption{Phase-locked loop}
        \label{fig:PLL}
    \end{figure}
   
The~{PLL} written in $dq$-coordinates is described by a second-order nonlinear model 
\begin{IEEEeqnarray}{rCl}\label{eq:PLLdq}
\dot{\mathbf{x}}\!\!&=&\!\mathbf{f}(\stateVarVec,\inputVarVec)\!=\!\!\begin{pmatrix}x_2-k_\mathrm{p}\left(u_1\sin{(x_1)}+u_2\cos{(x_1)}\right) \\
-k_\mathrm{i}\left(u_1\sin{(x_1)}+u_2\cos{(x_1)}\right)\end{pmatrix}\!,
\end{IEEEeqnarray}
where the states are the angle difference, and the integrator state, i.e., $\mathbf{x}=(\delta,{x}_\mathrm{I})^T$, and the inputs are the~$DQ$ voltage components,~i.e., $\mathbf{u}=(V_D,V_Q)$. The gains of the~PI controller are $k_\mathrm{p}, k_\mathrm{i}\in\mathbb{R}_{\geq0}$.
The right-hand side of~\eqref{eq:PLLdq} is described by the vector field~\mbox{$\mathbf{f}:\mathbb{R}^2\times\mathbb{R}^2 \to \mathbb{R}^2$}.

To transform the nonlinear model~\eqref{eq:PLLdq} into an~\ac{iMTI} model, the following smooth mapping is used  
\begin{IEEEeqnarray}{rCl}\label{eq:PLLtmap}
    \!\!\!\TrfVars\!\!:\!\!\begin{cases}
        \begin{aligned}
          \mathbb{R}^2\times \mathbb{R}^2&\to& \!\!\!\!\!\!\!\mathcal{M}:=\!\left\{\!\vec{z}\in\mathbb{R}^3,\vec{u}\in\mathbb{R}^2|\mathbf{g}(\vec{z})\!=\!0\right\}\!,\\
        \!\! (x_1,x_2,u_1,u_2)\!&\mapsto & \!\!\!\!\!\!\!\left(\cos{(x_1)},\sin{(x_1)},x_3,u_1,u_2 \right)\!,
    \end{aligned}
\end{cases}
\end{IEEEeqnarray}
where~\eqref{eq:PLL_dq_constraints}
constraints~$\mathcal{M}$. The new variables are\footnote{The time dependence is shown here to help the reader.}
\begin{IEEEeqnarray}{rCl} \label{eq:PLLnewVars}
 \MstateVarVec(t)&=&\begin{pmatrix}\cos{(x_1(t))},& \sin{(x_1(t))},&x_2(t)\end{pmatrix}^\T\,,
\end{IEEEeqnarray}
where for ease of notation the variables of the nonlinear model are set into~\mbox{$\bm{\kappa}(t)=(\stateVarVec(t)^{\T},\inputVarVec(t)^{\T})^{\T}$}, and the new variables in~\mbox{$\bm{\zeta}=(\MstateVarVec^{\T},\inputVarVec(t)^{\T})^{\T}$}.
Now~\eqref{eq:PLLnewVars} can be written as
\begin{IEEEeqnarray*}{rCl}
    \bm{\zeta}(t)=\mathbf{p}(\bm{\kappa}(t))\,.      
\end{IEEEeqnarray*}
To get the model in the new coordinates,~$\bm{\zeta}(t)$ is derivated with respect to time, where the chain rule needs to be considered
\begin{IEEEeqnarray*}{rCl}
      \frac{\mathrm{d}\bm{\zeta}(t)}{\mathrm{d}t}\!\!=\!\frac{\mathrm{d}\mathbf{p}(\bm{\kappa})}{\mathrm{d}\bm{\kappa}(t)}\frac{\mathrm{d} \bm{\kappa}(t)}{\mathrm{d} t}\!=\!\!\begin{pmatrix}
        -\sin{(\stateVar_1(t))}& 0 & 0 & 0\\
        \cos{(\stateVar_1(t))}& 0 & 0 & 0\\
        0& 1 & 0 & 0\\
        0& 0 & 1 & 0\\
        0& 0 & 0 & 1
    \end{pmatrix}\!\!
    \begin{pmatrix}
        \dot{\stateVar}_1(t)\\
        \dot{\stateVar}_2(t)\\
        0 \\
        0
    \end{pmatrix},
\end{IEEEeqnarray*}
which are expressed by the new variable yielding the  polynomial model
\begin{IEEEeqnarray}{rCl} \label{eq:PLLdq_mti_poly}
    \IEEEyesnumber
    \dot{z}_1&=&-z_2\left(z_3-k_\mathrm{p}\left(u_1 z_2+u_2z_1\right)\right)\,,\IEEEyessubnumber\\
        \dot{z}_2&=&z_1\left(z_3-k_\mathrm{p}\left(u_1z_2+u_2 z_1\right) \right)\,, \IEEEyessubnumber\\
        \dot{z}_3&=& -k_\mathrm{i}\left(u_1z_2+u_2z_1\right)\,.\IEEEyessubnumber
\end{IEEEeqnarray}
Two additional algebraic variables are introduced giving the~\ac{iMTI} model of the~\ac{PLL}

\begin{IEEEeqnarray}{rCl} \label{eq:PLLdq_mti}
    \vec{0}=
    \begin{pmatrix}
        \dot{z}_1+\left(z_3-k_\mathrm{p}\left(u_1 z_2+u_2z_1\right)\right)\auxVar_2\\
        \dot{z}_2-\left(z_3-k_\mathrm{p}\left(u_1z_2+u_2 z_1\right) \right) \auxVar_1\\
        \dot{z}_3+k_\mathrm{i}\left(u_1z_2+u_2z_1\right)\\
        z_1-\auxVar_1\\
        z_2-\auxVar_2
    \end{pmatrix}\,,
\end{IEEEeqnarray}
where the right-hand side is the implicit multilinear function~$\mathbf{h}:\mathbb{R}^3 \times\mathbb{R}^3\times \mathbb{R}^2 \times \mathbb{R}^2 \to \mathbb{R}^5$.

In the Appendix~\ref{appx:PLLlieblack}, it is shown that~\eqref{eq:PLLtmap} is a Lie-Bläcklund transformation. 
\end{example}

\subsection{On the Existence of a Solution and Numerical Treatment}
Contrary to the~\acp{ODE}, where there always exists a unique solution for an initial value if the right-hand side is smooth, the solution of general nonlinear~\acp{DAE} may not be straightforward~\cite{Kunkel.2006}. 
However, since the~\ac{iMTI} model of the power system is derived from a nonlinear~\ac{ODE} model, it is assumed that there exists a unique solution for the multilinear counterpart.
For the simulation of the~\ac{iMTI} model, the provision of the correct initial conditions that fulfill the algebraic constraints for the~\ac{iMTI} model is crucial to obtain a unique solution.

%% file: content/sec_efficient_linearization.tex
\section{Efficient Linearization of Implicit Multilinear Models} \label{sec:effLin}
The multilinear formulation of the power system components allow for an efficient computation of the Jacobian~\cite{Kaufmann.2023b}. Therefore, a linear model can be easily obtained, which is outlined in the following section. 

\subsection{Efficient Computation of the Jacobian}

The stucture of the~\ac{CPN1} format as a factorized multilinear polynomial allows for an efficient computation of the Jacobian matrix. 
The examples in~\cite{Kaufmann.2023b} showed a gain of computational performance by two orders of magnitude. 
This paper extends the method to~\ac{iMTI} models by calculating the partial derivates of~\eqref{eq:iMTIcp} with respect to all variables contained in the signal vector~$\vec{v}$.
Taking the partial derivative with respect to the~$i$-th variable of the~\ac{iMTI} model~\eqref{eq:iMTI_CPN1}, means computing the partial derivative of the structured contracted product~\eqref{eq:structuredCP} 
\begin{IEEEeqnarray}{rCl}
  && \left.\frac{\partial\vec{s}_r(\variableVector)}{\partial \variableVector_i}\right|_{\variableVector=\bar{\variableVector}}=\structureMatrix_{ir} \displaystyle\prod_{\substack{j=1 \\ j \neq i}}^{N_\mathrm{v}}\left(1-\!|\structureMatrix_{jr}|\!+\!\structureMatrix_{jr}\!\bar{\variableVector}_j\right)\,.
\label{eq:partialDeriv}
\end{IEEEeqnarray}
From~\eqref{eq:partialDeriv}, it can be seen that $i$-th term in~$\structuredCP(\bar{\variableVector})$ needs to be canceled out and the factor~$\structureMatrix_{ir}$ remains.
This structure is used to construct a numerically efficient procedure to calculate the Jacobian. 
Firstly, by computing a matrix~$\mathbf{D}\in\mathbb{R}^{R\times N_\mathrm{v}}$ with elements 
\begin{IEEEeqnarray*}{rCl}
  \mathbf{D}_{ir}\!=\!\begin{cases}
   \frac{\structureMatrix_{ir}}{1-|\structureMatrix_{ir}|\!+\!\structureMatrix_{ir}\bar{\variableVector}_i} & \text{iff } \parbox[t]{4cm}{$\structuredCP_r(\bar{\variableVector}) \neq 0,$\\$ \forall\, r=1,\ldots,R$\,,} \\
   \structureMatrix_{ir} & \text{iff }\parbox[t]{4cm}{$\left(1-\!|\structureMatrix_{ir}|\!+\!\structureMatrix_{ir}\bar{\variableVector}_i\right)=0,$} \\
   0 & \text{else}\,,
   \end{cases}
\end{IEEEeqnarray*}
and secondly, by evaluating the structured contracted at the operating point~$\bar{\variableVector}$
\begin{IEEEeqnarray*}{rCl}
  \structuredCP_{r}(\bar{\variableVector})\!=\!\begin{cases}
   \prod_{\substack{j=1}}^{N_\mathrm{v}}\left(1-\!|\structureMatrix_{jr}|\!+\!\structureMatrix_{jr}\!\bar{\variableVector}_j\right) & \text{iff } \parbox[t]{4cm}{$\structuredCP_r(\bar{\variableVector}) \neq 0,$\\$ \forall\, r=1,\ldots,R$\,,} \\
 \prod_{\substack{j=1 \\ j \neq i}}^{N_\mathrm{v}}\left(1-\!|\structureMatrix_{jr}|\!+\!\structureMatrix_{jr}\!\bar{\variableVector}_j\right) &\parbox[t]{4cm}{\text{iff }$ \\ \left(1-\!|\structureMatrix_{ir}|\!+\!\structureMatrix_{ir}\bar{\variableVector}_i\right)=0,$} \\
   0 & \text{else}\,. 
   \end{cases}
\end{IEEEeqnarray*}
Then the calculation of the Jacobian~\eqref{eq:partialDeriv} becomes a simple matrix multiplication
\begin{IEEEeqnarray}{rCl}\label{eq:jacobian}
  \mathbf{J}=\paramMatrix(\mathbf{D}\boxdot \structuredCP(\bar{\variableVector}) )^{\T}\in\mathbb{R}^{N_\mathrm{\phi}\times N_{v}}\,,
\end{IEEEeqnarray}
where the~\textit{broadcast product}~\cite{Harris.2020}, often used in scientific computing, e.g.,~\textsc{Matlab}, is denoted by $\boxdot$~\cite{Matsui.2024}. In this case, the elements of $\structuredCP(\bar{\variableVector})$ are duplicated to match the dimensions of~$\mathbf{D}$ allowing for an element-wise multiplication, i.e., the Hadamard product. 
More aspects for the implementation can be found in~\cite{Kaufmann.2023b}, which are similar to the case of~\ac{eMTI} models.

\subsection{Linear Descriptor Systems}\label{subsec:ldss}
 
\begin{definition}
A continuous-time~\ac{LDSS} is described by~\cite{Khalil.2015}
\begin{IEEEeqnarray}{rCl}
  \label{eq:LDSS}
  \mat{E}\begin{pmatrix}\dot{\MstateVarVec}\\ \dot{\auxVarVec}\end{pmatrix}&=&\mat{A}\begin{pmatrix}{\MstateVarVec}\\ {\auxVarVec}\end{pmatrix}+\mat{B}\vec{u}\,,\label{eq:LDSS_state}
\end{IEEEeqnarray}
where~$\MstateVarVec\in\mathbb{R}^{n}$, and~\mbox{$\auxVarVec\in\mathbb{R}^{q}$}, from the multilinear model, form the state vector of the~\ac{LDSS} model.  
The system matrix is denoted by~\mbox{$\mat{A}\in\mathbb{R}^{(n+q)\times (n+q)}$}, with the descriptor matrix~$\mat{E}\in \mathbb{R}^{(n+q)\times (n+q)}$ having the same dimensions, and~$m~\in \mathbb{N}$ inputs of the system are given by the input matrix~\mbox{$\mat{B}\in\mathbb{R}^{(n+q)\times m}$}. 
\end{definition}

To obtain an~\ac{LDSS}~\eqref{eq:LDSS} from an~\ac{iMTI} model, at the equilibrium point, meaning $\dot{\MstateVarVec}=\vec{0}$, and $\dot{\auxVarVec}=\vec{0}$~, 
 the matrices are extracted from the Jacobian~\eqref{eq:jacobian}
\begin{IEEEeqnarray}{rCl}
\label{eq:LSDSSmatrices}
  \,\mat{E}=- \left. \begin{pmatrix}
    \frac{\partial \vec{h}(\variableVector)}{\partial \dot{\MstateVarVec}} &\frac{\partial \vec{h}(\variableVector)}{\partial \dot{\auxVarVec}}
    \end{pmatrix} \right\rvert_{\substack{\mathbf{v}=\bar{\mathbf{v}}}} \,,
  &&
  \\
  \,\mat{A}= \left. \begin{pmatrix}
   
    \frac{\partial \vec{h}(\variableVector)}{\partial {\MstateVarVec}} &\frac{\partial \vec{h}(\variableVector)}{\partial {\auxVarVec}}
      \end{pmatrix} \right\rvert_{\mathbf{v}=\bar{\mathbf{v}}} \,,
   & &
   \,\mat{B}= \left. \begin{pmatrix}
       \frac{\partial \vec{h}(\variableVector)}{\partial {\vec{u}}} 
        \end{pmatrix} \right\rvert_{\mathbf{v}=\bar{\mathbf{v}}} \,,
   \IEEEnonumber 
\end{IEEEeqnarray}
where all variables are in $\variableVector=\left(\dot{\MstateVarVec}^\T,\,\dot{\auxVarVec}^\T,\,\newVarVec^\T,\,\auxVarVec^\T,\,\inputVarVec^\T\right)^\T$. The minus sign results from the implicit notation in~\eqref{eq:impFcn} and rearranging the descriptor matrix~$\mathbf{E}$ to the left-hand side.

With the obtained matrices, the linearized~\ac{iMTI} model can be written as an~\ac{LDSS}. 
Note that the coordinate system is moved into the equilibrium point,~i.e.,
\begin{equation*}
\mat{E}\begin{pmatrix}\dot{\MstateVarVec}\\ \dot{\auxVarVec}\end{pmatrix}=\mat{A}\begin{pmatrix}\MstateVarVec-\bar{\MstateVarVec}\\ \auxVarVec-\bar{\auxVarVec}\end{pmatrix}+\mat{B}\left(\vec{u}-\bar{\vec{u}}\right)\,,
\end{equation*} for instance for~\eqref{eq:LDSS_state}.

If~$\det{(\mat{E})}\neq0$, 
it can be inverted and~\eqref{eq:LDSS_state} can be written explicitly as 
\begin{IEEEeqnarray}{rCl}
\begin{pmatrix}\dot{\MstateVarVec}\\ \dot{\auxVarVec}\end{pmatrix}&=&\mat{E}^{-1}\mat{A}\begin{pmatrix}{\Delta\MstateVarVec}\\ {\Delta \auxVarVec}\end{pmatrix}+\mat{E}^{-1}\mat{B}\vec{u}\,.
\end{IEEEeqnarray}

If the~\ac{iMTI} model has a low number of factors, then it can be advantageous to store the~\ac{LDSS} model in the~\ac{CPN1} format. Furthermore, affine models that occur when the state derivatives are unequal to zero and a Taylor approximation is used, can also be modeled in the~\ac{CPN1} format. 
\begin{example}[PLL - continued]   
  \label{ex:PLL_dss}
  Linearizing the \ac{iMTI} model of the~\ac{PLL} by the computation of the Jacobian, and extracting the matrices using~\eqref{eq:LSDSSmatrices}, yields the following linear descriptor model   
  \setlength{\arraycolsep}{3pt}

  \begin{IEEEeqnarray*}{lCl}
    \mathbf{E}=\begin{pmatrix}
      -1 &0 & 0 & 0 & 0\\
      0 &-1 & 0 & 0 & 0\\
      0 &0 & -1 & 0 & 0\\
      0 &0 &0 & 0 & 0\\
      0 &0 &0 & 0 & 0
    \end{pmatrix}\,,
  \end{IEEEeqnarray*}
      \begin{IEEEeqnarray*}{lCl}
    \mathbf{A}=\\
    \left.\begin{pmatrix}
     -k_\mathrm{p}\auxVar_{2}u_{2} &  -k_\mathrm{p}\auxVar_{2}u_{1} & \auxVar_{2} & 0 &  \newVar_{3}-g(\newVarVec)\\
        k_\mathrm{p}\auxVar_{1}u_{2} &  k_\mathrm{p}\,\auxVar_{1}u_{1} &- \auxVar_{1} &   g(\newVarVec) -\newVar_{3}& 0\\
        k_\mathrm{I}u_{2} & k_\mathrm{I}u_{1} & 0 & 0 & 0\\
         1 & 0 & 0 & -1 & 0\\ 
         0 & 1 & 0 & 0 & -1 
    \end{pmatrix}\right|_{\substack{\newVarVec=\bar{\newVarVec},\\ \auxVarVec=\bar{\auxVarVec},\\ \inputVarVec=\bar{\inputVarVec}}}\,,\\   
    \mathrm{where }\,\,g(\newVarVec)=k_\mathrm{p}\left(\newVar_{1}\,u_{2}+\newVar_{2}\,u_{1}\right)\,,\,\text{and}\\
    \\
    \mathbf{B}=
    \left.\begin{pmatrix}
      -k_\mathrm{P}\newVar_{2}\,\auxVar_{2} & -k_\mathrm{P}\,\newVar_{1}\,\auxVar_{2}\\
       k_\mathrm{P}\,\newVar_{2}\,\auxVar_{1} & k_\mathrm{P}\,\newVar_{1}\,\auxVar_{1}\\
        k_\mathrm{I}\,\newVar_{2} & k_\mathrm{I}\,\newVar_{1}\\
         0 & 0\\
          0 & 0
    \end{pmatrix}\right|_{\substack{\newVarVec=\bar{\newVarVec},\\ \auxVarVec=\bar{\auxVarVec},\\ \inputVarVec=\bar{\inputVarVec}}}\,.
  \end{IEEEeqnarray*} 
\end{example}

%% file: content/sec_gen_eig.tex
\section{Small-signal Stability Analysis by Generalized Eigenvalues}\label{sec:GEP}
The stability of operating points of nonlinear power systems in the sense of Lyapunov can be assessed locally by checking the eigenvalues of the linearized model~\mbox{\cite[p. 49]{Milano.2020}}. 
Often these are linear eigenvalue problems, while for linear descriptor systems, the generalized eigenvalue problem applies~\cite{Milano.2017}. Therefore, the stability of the operating points of~\ac{iMTI} models can be assessed by analyzing the generalized eigenvalue of the obtained~\ac{LDSS}, which is described in this section. 

The~\ac{GEP} is~\cite{Milano.2017}
\begin{IEEEeqnarray}{rCl}
  \IEEEyesnumber
   \mat{A}\vec{v}=\lambda\mat{E}\vec{v}\,,\IEEEyessubnumber \label{eq:gep_right} \\
   \vec{w}\mat{A}=\lambda \vec{w} \mat{E}\,, \IEEEyessubnumber\label{eq:gep_left} \
\end{IEEEeqnarray}
where~$\lambda\in\mathbb{C}$ is the eigenvalue,~$\vec{v}\in\mathbb{C}^{(n+q)\times 1}$ the corresponding right eigenvector, and~$\vec{w}\in\mathbb{C}^{1\times(n+q)}$ the left eigenvector. 
If~$\mat{E}$ is equal to the identity matrix, meaning there are no algebraic equations,~the~\ac{GEP} reduces to the common eigenvalue problem.
The matrix pair~$(\mat{E},\mat{A})$ form a matrix pencil. 
A matrix pencil is a matrix-valued polynomial function, though for \ac{LDSS}, it is a linear matrix pencil~$\lambda\mat{E}-\mat{A}$, which is a family of matrices parameterized by a complex number~\mbox{$\lambda\in\mathbb{C}$~\cite[p. 66]{Milano.2020}\cite{Gantmacher.1959}}. 
Due to the construction of the~\ac{LDSS},~i.e.,~$\mat{E}$ and~$\mat{A}$ are square matrices, the matrix pencil of the derived~\ac{LDSS} from the~\ac{iMTI} model has a characteristic polynomial. 
The generalized eigenvalues are the roots of the characteristic polynomial of the matrix pencil~$(\mat{A},~\mat{E})$. 
The order of the characteristic polynomial are the~$k$~finite eigenvalues of the regular pencil. 
The regular pencil can have~$l$ infinite eigenvalues, where~\mbox{$k+l=n+q$}. 
Considering the reciprocal generalized eigenvalue problem~\mbox{$\lambda^{-1} \mat{A}\vec{v}=\mat{E}\vec{v}$}, if~$\mat{E}$ is singular, there exists a null vector~$\vec{v}$, such that~\mbox{$\mat{E}\vec{v}=\vec{0}\in\mathbb{R}^{(n+q)\times 1}$}.
Thus,~\mbox{$\lambda^{-1}\mat{A}\vec{v}=\vec{0}\in\mathbb{R}^{(n+q)\times 1}$}, so that $\lambda^{-1}=0$, corresponing to $\lambda \rightarrow \infty$~\cite{Milano.2017}. 
This case is demonstrated in the following.
\begin{example}Consider the following autonomous descriptor model with two states
  \begin{IEEEeqnarray*}{rCl}
    \begin{pmatrix}
      1 & 0 \\ 0 & 0
    \end{pmatrix}
    \begin{pmatrix}
      \dot{\MstateVar}_1 \\ \dot{\MstateVar}_2 
    \end{pmatrix}
    =
    \begin{pmatrix}
      0 & 1\\
      1 & 1
    \end{pmatrix}
    \begin{pmatrix}
      \MstateVar_1 \\ \MstateVar_2 
    \end{pmatrix} \,.   
  \end{IEEEeqnarray*}
The descriptor matrix~$\mat{E}$ is singular since~$\det(\mat{E})=0$, meaning there is one right eigenvector that is in the nullspace of~$\mat{E}$. 

Computing the generalized eigenvalues and eigenvectors using Matlab's~\texttt{eig} command yields~$\lambda_1=-1$, and~\mbox{$\lambda_2=\texttt{Inf}$}, and
\begin{IEEEeqnarray*}{rCl}
  \mat{V}=\begin{pmatrix}
    1 & 0 \\
   -1 & 1
  \end{pmatrix},\,
  \mat{W}=\begin{pmatrix}
    1 & 0 \\
   -1 & 1
  \end{pmatrix}\,,
\end{IEEEeqnarray*}
where the second right eigenvector gives the zero vector
\begin{IEEEeqnarray*}{rCl}
      \begin{pmatrix}
       1 & 0 \\ 0 & 0
      \end{pmatrix}
        \begin{pmatrix}
          0 \\
          1
        \end{pmatrix}
    =
    \begin{pmatrix}
      0 \\
      0
    \end{pmatrix}\,,
\end{IEEEeqnarray*}
corresponding to the infinite eigenvalue, i.e., the null space of~$\mathbf{E}$ .
\end{example}

The generalized eigenvalues are used to assess if an equilibrium point of the~\ac{iMTI} model is unstable, according to the following definition. 
\begin{definition}~\cite{Milano.2020}
  The equilibrium point~$(\bar{\variableVector})$ of the~\ac{iMTI} model is locally unstable if at least one of the generalized eigenvalues of the~\ac{LDSS} are on the right-half of the complex plane,~i.e.,~$\Re{(\lambda)}>0$. 
\end{definition}
With this definition, the small-signal stability of an equilibrium point of the~\ac{PLL} is assessed in the following using the generalized eigenvalues.  
\begin{example}[PLL - continued]
  The generalized eigenvalues of the obtained linear descriptor model of the~\ac{PLL} given in Example~\ref{ex:PLL_dss} are evaluated in the following. 
  The rank of the descriptor matrix~$\mathrm{rank}(\mathbf{E})=3$, therefore, two generalized eigenvalues will be infinite. 
  The system matrix is of~$\mathrm{rank}(\mathbf{A})=4$, thus there will be one zero eigenvalue. 
  The~\ac{PLL} has the proportional and integral gains~\mbox{$k_\mathrm{P}=\SI{0.5}{\radian \per \volt \per \second}$},~and~\mbox{$k_\mathrm{I}=\SI{9}{\radian \per \volt \per \second^2}$}, respectively.
  After an angle step of $\delta=\SI{4}{\degree}$, the~\ac{iMTI} model reaches the new equilibrium 
  \begin{IEEEeqnarray*}{lll}
     \bar{\newVar}_1=\bar{\newVar}_4=0.9976,&\,\bar{\newVar}_2=\bar{\newVar}_5=-0.0698,&\,\bar{\newVar}_3=\SI{0}{\radian \per \second}\,,\\
     \bar{\inputVar}_1=\SI{325.2059}{\volt}, &\, \bar{\inputVar}_2=\SI{22.7406}{\volt} \,.& 
  \end{IEEEeqnarray*} 
At this equilibrium point, the~\ac{LDSS} has the generalized eigenvalues
\begin{IEEEeqnarray*}{llll}
  \lambda_1=-20.6046\,,&\,  \lambda_2=-142.3954\,, &\, \lambda_3=0\,, &\, \lambda_{4,5}=\texttt{Inf}\,,
\end{IEEEeqnarray*}
which means that the equilibrium point is stable.

For a comparison, the Jacobian of the nonlinear model of the~\ac{PLL}~\eqref{eq:PLLdq} is computed with respect to~$x_1, x_2$
\begin{IEEEeqnarray*}{lCl}
 \mathbf{A}=\begin{pmatrix}
    -k_\mathrm{P}\left(\inputVar_1\cos{(\stateVar_1)}-\inputVar_2\sin{(\stateVar_1)}\right) & 1\\
    -k_\mathrm{I}\left(\inputVar_1\cos{(\stateVar_1)}-\inputVar_2\sin{(\stateVar_1)}\right) &0
  \end{pmatrix}\,, 
\end{IEEEeqnarray*}
which yields the eigenvalues 
\begin{IEEEeqnarray*}{ll}
  \lambda_1=-142.3954\,,&\quad 
  \lambda_2=-20.6046\,. 
\end{IEEEeqnarray*}
Therefore, both models come to the same conclusion that the studied equilibrium point is stable.
\end{example}

%% file: content/sec_model_case_study.tex
\section{Power System Model}\label{sec:ps_ex}
\begin{figure}[b]
    \centering
    \includegraphics[width=\columnwidth]{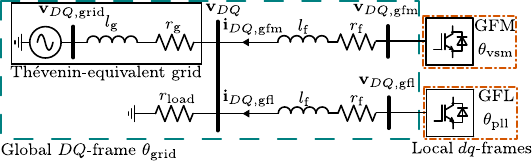}%
    \caption{Scheme of the essential system in the global frame~$\theta_\mathrm{grid}$ (dashed line), and the local frames of the converters (dashed-dotted line)}
    \label{fig:GFMgfl}
\end{figure}

To test the proposed methodology, an essential power system network that reflects the relevant dynamics of converter-dominated networks is constructed. 
It consists of a~\ac{GFM} converter, and a~\ac{GFL} converter, depicted in Fig.~\ref{fig:GFMgfl}, is constructed in the following section.
The models are implemented in~\textsc{Matlab}~\cite{MATLAB.2023} using the MTI-Toolbox~\cite{Lichtenberg.2024}, as well as in Simulink for a comparison. 

\subsection{System Description in $dq$-Coordinates}
Similar to other small-signal power system stability studies~\cite{Pogaku.2007,Markovic.2021}, the network is modeled in $dq$-coordinates.
Each device has its own rotating $dq$-frame, indicated by the lower case, while the network is modeled in a global rotating frame indicated by the upper case~$DQ$. 
This modeling approach helps in the linearization of the model because for small angle deviations between the global and the local frame, the trigonometric functions in the rotational matrices can be linearly approximated~\cite{Pogaku.2007}.  
Therefore, the rotational matrix~$\mathbf{R}(\theta_i,\theta_\mathrm{grid})\in \mathbb{R}^{2\times 2}$ is applied to the grid voltage and current, to rotate from~$DQ \rightarrow dq$ 
\begin{IEEEeqnarray}{rCl}
    \IEEEyesnumber \label{eq:rotationalMatrix}
    \mathbf{R}(\theta_i,\theta_\mathrm{grid})&=& \begin{pmatrix}  h_1(\theta_i,\theta_\mathrm{grid}) &   h_2(\theta_i,\theta_\mathrm{grid})  \\
        - h_2(\theta_i,\theta_\mathrm{grid})&   h_1(\theta_i,\theta_\mathrm{grid})
            \end{pmatrix} \\
        h_1(\theta_i,\theta_\mathrm{grid})&=&\cos{({\theta}_i-{\theta}_\mathrm{grid})}\IEEEnonumber \\ 
                                        &=&\cos{({\theta}_i)}  \cos{({\theta}_\mathrm{grid})} +\sin{({\theta}_i)}  \sin{({\theta}_\mathrm{grid})} \IEEEnonumber\\
        h_2(\theta_i,\theta_\mathrm{grid})&=&\sin{({\theta}_i-{\theta}_\mathrm{grid})} \IEEEnonumber\\ 
                &=&\sin{({\theta}_i)}  \cos{({\theta}_\mathrm{grid})} -\cos{({\theta}_i)}  \sin{({\theta}_\mathrm{grid})}\,, \IEEEnonumber     
    \end{IEEEeqnarray}
where~$\theta_i\in\mathbb{R}$ is the phase angle of the local device, and $\theta_\mathrm{grid}\in\mathbb{R}$ is the angle of the global~\ac{SRF}. To rotate from~$dq\rightarrow DQ$, the inverse rotation $\mathbf{R}^{-1}(\theta_i,\theta_\mathrm{grid})$ is used.

\subsection{Network Model}\label{subsec:model_network}
The network operates at high voltage~\SI{230}{\kilo \volt} with~\SI{50}{\hertz} and it is modeled as a Th\'evenin equivalent with a~\mbox{\ac{SCR}} of~$5$, a base apparent power of~\SI{100}{\mega \volt \ampere}, and a $X/R$-ratio of~$10$. 

For a general $RL$~circuit between two nodes~$k$ and~$l$, with voltages $\vec{v}_{DQ,k},\vec{v}_{DQ,l}\in\mathbb{R}_{\geq0}^2$, and with the current~$\vec{i}_{DQ,kl}\in\mathbb{R}^2$ flowing from~$k$ to~$l$, the current is described by
\begin{IEEEeqnarray}{rCl}
    \dot{\vec{i}}_{DQ,kl}&=&\left(-\frac{r_{kl}}{l_{kl}}-\omega_g\bm{\mathcal{J}}_2\right){\vec{i}}_{DQ,kl}\IEEEnonumber\\
    &&+\frac{1}{l_{kl}}\left(\vec{v}_{DQ,k}-\vec{v}_{DQ,l} \right)\,,    
    \label{eq:RL}
\end{IEEEeqnarray}
where~$r_{kl}, l_{kl}\in\mathbb{R}_{>0}$ are the resistance and inductance values of the section~$kl$, the angular grid frequency~\mbox{$\omega_\mathrm{g}\in \mathbb{R}_{>0}$}, and for the cross-coupling in $dq$-coordinates
\renewcommand{\arraystretch}{0.3}~\mbox{
$\mathbf{\mathcal{J}}_2=\textstyle\begin{pmatrix}
    0 & -1\\ 1 & 0
\end{pmatrix}$}. If the grid frequency is either a state or an output of the system, the $RL$-model is multilinear. The~$RL$ circuit model~\eqref{eq:RL} is used for the Th\'evenin equivalent grid, and for the filter equations of the converters. 

The network has a resistive load~$r_\mathrm{load}\in\mathbb{R}_{>0}$ which yields the voltage 
\begin{IEEEeqnarray}{rCl}
    \vec{v}_{DQ}=r_\mathrm{load} \Delta \vec{i}_{DQ},
    \label{eq:load}
\end{IEEEeqnarray}
where~$\Delta \vec{i}_{DQ}$ is the difference between the currents flowing into the node. 

\subsection{Grid-Forming Converter}\label{subsec:casestudy_GFM}
\begin{figure}
    \centering
    \includegraphics[width=\columnwidth]{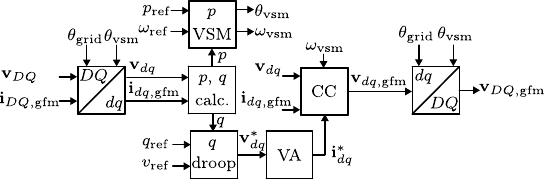}%
    \caption{Cascaded control loop structure of the grid-forming converter}
    \label{fig:GFMcascaded}
\end{figure}

 A~{\ac{VSM}}~\cite{Zhong.2011} is used for the active power control of the~\ac{GFM}~converter, and a filtered voltage droop control for the reactive power control~\cite{Markovic.2021}.
It was shown that different~\ac{GFM} control strategies behave like a droop-based controller in a small-signal regime~\cite{Markovic.2021,Gros.2019}. 
The implemented~\ac{GFM} controller uses a cascaded control structure with a~\ac{VA}, and a~\ac{CC}-loop, as depicted in~Fig.~\ref{fig:GFMcascaded}. 

After rotating the voltage~$\mathbf{v}_{DQ}\in\mathbb{R}^2$, and the current~$\mathbf{i}_{DQ}\in\mathbb{R}^2$, to the local coordinates using~\eqref{eq:rotationalMatrix}, i.e,~$\mathbf{v}_{dq}\in\mathbb{R}^2$, and~$\mathbf{i}_{dq}\in\mathbb{R}^2$, respectively, the active power~$p\in\mathbb{R}$, and the reactive power~$q\in\mathbb{R}$ are calculated. 

\subsubsection{Active Power Control}
The active power control is based on the~\ac{VSM} approach, called synchronverter~\cite{Zhong.2011}, that tries to mimic the frequency dynamics of a~\mbox{\ac{SG}~\cite{Zhong.2011,Cardozo.2024}}
\begin{IEEEeqnarray}{rCl}
    \IEEEyesnumber \label{eq:VSM}
    \dot{\theta}_\mathrm{vsm}&=&\omega_\mathrm{b}\omega_\mathrm{vsm}\,, \IEEEyessubnumber \label{eq:VSMangle}\\
       2 h \dot{\omega}_\mathrm{vsm}&=&(p_\mathrm{ref} -\frac{p}{s_\mathrm{b}})+k_\mathrm{d}(\omega_\mathrm{vsm}-\omega_\mathrm{ref})\,,
    \IEEEyessubnumber \label{eq:VSMfrequency}
\end{IEEEeqnarray}
where the controller uses the~$\si{\pu}$ quantities for the active power reference~$p_\mathrm{ref}~\in(-1,1)$, and the angular frequency reference~$\omega_\mathrm{ref}\in[0,1)$, the base angular frequency~\mbox{$\omega_\mathrm{b}\in \mathbb{R}_{>0}$}, and the base apparent power~\mbox{$s_\mathrm{b}\in \mathbb{R}_{>0}$}. 
The angular frequency dynamics~$\omega_\mathrm{VSM}\in\mathbb{R}$ depend on the virtual inertia's constant~$h\in \mathbb{R}_{\geq0}$, that can be chosen to meet the system inertia requirement~\cite[Sec. V-d]{Cardozo.2024}, and the choice of the (virtual) damping $k_\mathrm{d}\in \mathbb{R}_{\geq0}$, which is a design choice and~\cite[Sec. V-d]{Cardozo.2024} provides a guideline.

The rotational matrix~\eqref{eq:rotationalMatrix} uses the angle~$\theta_\mathrm{VSM}$. Therefore, two additional states are introduced following the procedure outlined in Section~\ref{subsubsec:TrigFunc} 
\begin{IEEEeqnarray}{rCl}
    \IEEEyesnumber \label{eq:}
    \dot{z}_\mathrm{cos,vsm}&=&-\omega_\mathrm{b} \omega_\mathrm{vsm} z_\mathrm{sin,vsm}\IEEEyessubnumber,\\ 
    \dot{z}_\mathrm{sin,vsm}&=&\omega_\mathrm{b} \omega_\mathrm{vsm} z_\mathrm{cos,vsm}\IEEEyessubnumber,
\end{IEEEeqnarray}
where the constraint as in~\eqref{eq:trigConstraint} is not explicitly given here. 

\subsubsection{Reactive Power Control}
For the reactive power control, a droop control with a low-pass filter is used~\cite{Markovic.2021},
\begin{IEEEeqnarray}{rCl}
    \IEEEyesnumber
    \dot{\stateVar}_{q,\mathrm{filt}}&=&-\omega_f \stateVar_{q,\mathrm{filt}}+\omega_f q\,, \IEEEyessubnumber\\
    {v}_\mathrm{d}^*-v_\mathrm{ref}&=&-k_\mathrm{q}\left(\stateVar_{q,\mathrm{filt}}-q_\mathrm{ref}\right)\,,\IEEEyessubnumber
    \label{eq:reactive_power_droop}
\end{IEEEeqnarray}
where~$\stateVar_{q,\mathrm{filt}}~\in\mathbb{R}$ is the integrator state of the low-pass filter,~$\omega_f \in \mathbb{R}_{\geq0}$ is the cut-off frequency, and~$k_\mathrm{q}\in \mathbb{R}_{\geq0}$ is a proportional gain that emulates the droop characteristic of the~\ac{SG}. The reactive power reference~$q_\mathrm{ref}$ is given in $\si{\var}$, and $v_\mathrm{peak}\in\mathbb{R}_{>0}$ is the nominal peak line-to-ground voltage in~\si{\volt}, so that the generated reference value of the $d$-component~${v}_\mathrm{d}^*\in\mathbb{R}_{>0}$ is also in~\si{\volt}. Note that~$v_\mathrm{q,ref}=0$ for the voltage regulation. 

\subsubsection{Virtual Admittance}
A virtual admittance can reduce the instantaneous change in active power of the~\ac{GFM} converter to large disturbances by adjusting the current reference~$\mathbf{i}_\mathrm{dq}^*\in\mathbb{R}^2$. 
Lowering the instantaneous change in active power can help the converter to withstand larger disturbances, e.g., large phase jumps of~\SI{30}{\degree}, and remain connected~\cite[Sec. IV.-D]{Cardozo.2024}. A state-space representation of the virtual admittance is 
\begin{IEEEeqnarray}{rCl}
    \IEEEyesnumber
     \dot{\mathbf{x}}_{\mathrm{va},dq}&=&\left(-\frac{r_\mathrm{v}}{l_\mathrm{v}}+\omega_\mathrm{b} \bm{\mathcal{J}}_2\right)  {\mathbf{x}}_{\mathrm{va},dq} +  \mathbf{v}_\mathrm{dq,ref}\,, \IEEEyessubnumber\\
     \mathbf{i}_\mathrm{dq}^*&=&\frac{1}{l_\mathrm{v}}{\mathbf{x}}_{\mathrm{va},dq}\,,\IEEEyessubnumber
\end{IEEEeqnarray}
where the virtual inductance and the virtual resistance are~\mbox{$l_\mathrm{v},r_\mathrm{v}\in\mathbb{R}_{>0}$}, respectively. The integrator state of the virtual admittance is denoted by~$\mathbf{x}_\mathrm{va,dq}\in\mathbb{R}^2$. The cross-coupling between the $dq$-axis is indicated by~$\mathbf{\mathcal{J}}_2$. 
According to~\cite{Leon.2023}, the output inductance of~\ac{SG} is \SIrange{30}{50}{\percent} and is used as a reference for the range for the virtual inductor~$k_\mathrm{va}\in[0,1)$. 
The virtual resistor should be around one third of the virtual inductor~$r_\mathrm{v}=k_\mathrm{va}\frac{z_\mathrm{b}}{3}$ with~\mbox{$l_\mathrm{v}=k_\mathrm{va}{z_\mathrm{b}}/{(\omega_\mathrm{b})}$}~\cite{Leon.2023}.

\subsubsection{Current Control}
To follow the current reference from the virtual admittance, a common current control loop using a PI control is used. 
It is based on the internal model principle considering the converter filter~\cite{Harnefors.1998}. 
The multilinear model of the current control loop is
\begin{IEEEeqnarray}{rCl}
  \IEEEyesnumber  \label{eq:CCloop} 
    \dot{\mathbf{x}}_{i,dq}&=&k_{i,\mathrm{i}}\left(\mathbf{i}_\mathrm{dq,ref}-\mathbf{i}_\mathrm{dq}\IEEEyessubnumber\right)\,,\\
    \mathbf{v}_{dq,\mathrm{gfm}}&=&k_{i,\mathrm{p}}\left(\mathbf{i}_{dq,\mathrm{ref}}-\mathbf{i}_\mathrm{dq}\right)+k_{i,\mathrm{i}}{\mathbf{x}}_{I,dq}\IEEEnonumber\\
    &&+l_f \omega_\mathrm{b}\omega_\mathrm{vsm} \bm{\mathcal{J}}_2 \mathbf{i}_{dq}+\mathbf{v}_{dq}\,,\IEEEyessubnumber \label{eq:CCloopOutput}
\end{IEEEeqnarray}
where~$\mathbf{x}_{i,dq}\in\mathbb{R}^2$ denotes the integrator state, the gains of the PI controller are $k_{i,\mathrm{i}}, k_{i,\mathrm{p}}\in\mathbb{R}_{\geq0}$, and~$\mathbf{u}_{dq,\mathrm{ref}}\in\mathbb{R}^2$ is the reference voltage that is usually sent to the~pulse-width modulation signal generator to obtain the gating signal for the switches, but here considered as the output voltage of the converter.
 The inverse rotation of~\eqref{eq:rotationalMatrix} is applied to~\mbox{$\mathbf{v}_{dq,\mathrm{gfm}}$} to get the output voltage of the converter in the global coordinate frame. 
 The closed-loop response of the PI-controller is designed as a first-order response with~\mbox{$\tau_\mathrm{cc}=\SI{1}{\milli \second}$}, and~\mbox{$k_{i,\mathrm{p}}={l_f}/{\tau_\mathrm{cc}}$}, and~\mbox{$k_{i,\mathrm{i}}={r_f}/{\tau_\mathrm{cc}}$} for the gains~\cite{Yazdani.2010}.

\begin{table}[tb]
    \renewcommand{\arraystretch}{1.2}
    \caption{Converter Parameters }
    \begin{tabularx}{\columnwidth}{p{0.9cm} X p{0.9cm} X }
        \hline
        \textbf{Symbol} & \textbf{Value}  & \textbf{Symbol} & \textbf{Value} \\
        \hline
        $s_\mathrm{b}$ & \SI{100}{\mega\volt\ampere} &  $r_\mathrm{f}$ & \SI{0.02}{\pu} \\
        $l_\mathrm{f}$ & \SI{0.07}{\pu} & $h$ & \SI{1}{\second} \\
        $k_\mathrm{d}$ & $50$ & $\omega_{ref}$ & \SI{1}{\pu} \\
        $\omega_f$&\SI{20}{\radian}&$k_q$ & \SI{3.7559e-04}{\volt \per \var} \\
        $k_\mathrm{p,CC}$ & \SI{117.87}{\henry\per\second} &  $k_\mathrm{i,CC}$ &\SI{1058}{\ohm\per\second} \\
        $r_\mathrm{v}$ & \SI{0.30}{\ohm} & $l_\mathrm{v}$ & \SI{0.03}{\henry} \\
        $k_\mathrm{p,PLL}$ & \SI{1.065e-4}{\radian\per\volt\per\second} &  $k_\mathrm{I,PLL}$ & \SI{0.0011}{\radian\per\volt\per\second^2} \\
        $k_{\mathrm{p},pq}$ & \SI{10e-5}{}& $k_{\mathrm{i},pq}$ & \SI{1e-2}{\per\second} \\
        \hline
    \end{tabularx}
    \label{tab:convContParam}
\end{table}

\subsection{Grid-Following Converter}
\begin{figure}
    \centering
    \includegraphics[width=\columnwidth]{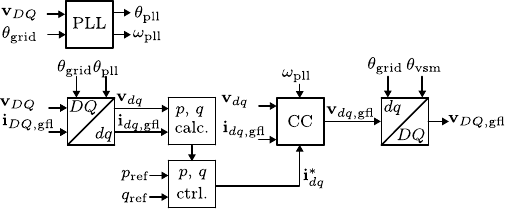}%
    \caption{Control block diagram of the grid-following converter}
    \label{fig:gfl}
\end{figure}

The cascaded controller structure of the~\ac{GFL}~converter is depicted Fig.~\ref{fig:gfl}. 
The~\ac{GFL} converter uses a~\ac{PLL} to synchronize with the grid voltage, as shown in~\eqref{eq:PLLdq_mti}, which is tuned following~\cite{Ch00}, where the gains are~$k_\mathrm{p,PLL}$, and~$k_\mathrm{i,PLL}$. 
A $pq$-controller is used to regulate the power exchange of the converter with the network by providing the current reference signal~$i_{dq}^*$. 
The active and reactive power are controlled by two separate PI-controllers with the gains~$k_{\mathrm{p},pq}$, and~$k_{\mathrm{i},pq}$. 
The current controller then uses~$i_{dq}^*$ to generate the voltage at the converter terminal~$\mathbf{v}_{dq,\mathrm{gfl}}$, and it has the same structure and tuning as in~\eqref{eq:CCloop}. 

\subsection{Complete Multilinear Model}
The tuning and sizing values of the~\ac{GFM} and~\ac{GFL} converter are listed in Table~\ref{tab:convContParam}, besides the parameter in Section~\ref{subsec:model_network} 
The total~\ac{iMTI} model has the following states of each subsystem, and the input vector for the whole model, as follows
\begin{IEEEeqnarray}{rCl}
    \IEEEyesnumber \label{eq:GFMss}
    \MstateVarVec_\mathrm{gfm}&=& (\omega_\mathrm{vsm},\,\theta_\mathrm{VSM},\,z_\mathrm{cos,vsm},\,z_\mathrm{sin,vsm},\, \stateVar_{q,\mathrm{filt}},\IEEEnonumber\\
    && \mathbf{x}_{\mathrm{va},dq},\, {\mathbf{x}}_{i,dq}
    )^{\T}\in\mathbb{R}^{9},\IEEEyessubnumber\\
      \MstateVarVec_\mathrm{gfl}&=& (z_\mathrm{cos,pll},\,z_\mathrm{sin,pll},\, {x}_\mathrm{i,pll},\theta_\mathrm{pll},\,{x}_{\mathrm{i},p},\,{x}_{\mathrm{i},q},\IEEEnonumber\\
    &&    {\mathbf{x}}_{\mathrm{i},dq} )^{\T}\in\mathbb{R}^{9},\IEEEyessubnumber\\
        \MstateVarVec_\mathrm{grid}&=& (\mathbf{i}_{DQ,\mathrm{grid}},\,\mathbf{i}_{DQ,\mathrm{gfm}},\,\mathbf{i}_{DQ,\mathrm{gfl}},\IEEEnonumber\\
        &&\,z_\mathrm{cos,grid},\,z_\mathrm{sin,grid})^{\T}\in\mathbb{R}^{8},\IEEEyessubnumber\\   
        \vec{u}_\mathrm{total}&=&
        (p_\mathrm{ref,gfm},q_\mathrm{ref,gfm},\omega_\mathrm{ref},v_\mathrm{ref},\IEEEnonumber\\
       && p_\mathrm{ref,gfl},q_\mathrm{ref,gfl})^{\T}\in\mathbb{R}^6, \IEEEyessubnumber
 \end{IEEEeqnarray}
where $\MstateVarVec_\mathrm{total}=\left(\MstateVarVec_\mathrm{gfm}^\T,\MstateVarVec_\mathrm{gfl}^\T,\MstateVarVec_\mathrm{grid}^\T\right)^\T$. 
The output and intermediate variables are not explicitly stated here.
The~\ac{iMTI} model of the total network in~\ac{CPN1} has $165$ factors, with~$58$ equations. 

There is a special case for the calculation of the Jacobian because of the nonlinear change of coordinates to represent the trigonometric functions, as in the~\ac{PLL} in Example~\ref{ex:PLL1}. 
If the partial derivative of the~\ac{iMTI} needs to be taken with respect to a variable of the original model, which are in this example angles~$\theta$, then the chain rule applies 
\begin{IEEEeqnarray*}{rCl}
   \left.\frac{\partial\vec{h}(\variableVector)}{\partial \stateVarVec_i}\right|_{\variableVector=\bar{\variableVector}}&=& \left.\frac{\partial\vec{h}(\variableVector)}{\partial \variableVector_i}\right|_{\variableVector=\bar{\variableVector}} \left.\frac{\partial\vec{f}(\vec{\kappa})}{\partial \vec{\kappa}_i}\right|_{\vec{\kappa}=\bar{\vec{\kappa}}}\,,\\
\end{IEEEeqnarray*}
which generalizes to 
\begin{IEEEeqnarray}{rCl}    
    \mathbf{J}_{\TrfVars\circ \vec{f}}(\bar{\mathbf{x}},\bar{\mathbf{u}})=\mathbf{J}_{\TrfVars}(\mathbf{f}(\bar{\mathbf{x}},\bar{\mathbf{u}}))\,\mathbf{J}_{\mathbf{f}}(\bar{\mathbf{x}},\bar{\mathbf{u}}) \,.      
  \end{IEEEeqnarray}
In this paper, only the first partial derivative is considered. 
Since the computation of the Jacobian of the~\ac{iMTI} model is an analytical computation, and the derivatives of the nonlinear functions with respect to time can be precomputed, the Jacobian only needs to be computed once. 
The operating point is then inserted, maintaining an efficient approach for the linearization.

%% file: content/sec_results_case_study.tex
\section{Simulation Results, Small-signal Analysis, and Discussion}\label{sec:results}
\begin{figure}
    \centering
    \includegraphics{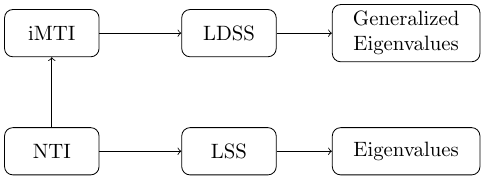}
    \caption{Comparison of the small-signal stability analysis methods using~\ac{iMTI} modeling and~\ac{NTI} modeling}
    \label{fig:methodFlow}
\end{figure}

The proposed methodology to model power systems and performing small-signal stability analysis using multilinear models is tested for different scenarios. 
Fig.~\ref{fig:methodFlow} illustrates the comparison of the small-signal stability analysis methods using~\ac{iMTI} modeling and~\mbox{\ac{NTI}} modeling.
Firstly, the results of the time-domain simulations of the nonlinear, multilinear, and the obtained linear model are compared. 
Then the small-signal stability analysis is performed where the generalized eigenvalues of the obtained descriptor state-space model are compared with the eigenvalues of the linear state-space model approximating the nonlinear model at the equilibrium point. 
The computation time required for the linearization will be compared at the end.

\subsection{Comparison of the Simulation Results}
The nonlinear model of the 3-bus network was implemented in Simulink of the Matlab 2023b version. 
The~\ac{iMTI} model is initialized from the time-domain simulation of the~\ac{NTI}~model. 
Note that the algebraic constraint~\eqref{eq:trigConstraint} is used in addition to compute consistent initial conditions for the~\ac{iMTI} model.
Similarly, the equilibrium point is obtained from the~\ac{NTI}~model, at which the linearization of the~\ac{iMTI} model is performed using~\eqref{eq:jacobian}, and the~\ac{LDSS} model is constructed using~\eqref{eq:LDSS_state}.
The equilibrium point is chosen by inspection.
 
The~\texttt{ode23} algorithm for the solution of initial value problems in Simulink~\cite{MATLAB.2023} is used for the~\ac{NTI}~model, which is a single-step solver based on the explicit Runge-Kutta (2,3) pair formula and suitable for moderately stiff problems with crude tolerances. 
The~\ac{iMTI} model is simulated using the~\texttt{MTI Toolbox}~\cite{Lichtenberg.2024}, which uses the~\texttt{ode15i} solver~\cite{MATLAB.2023}, which is an implicit variable-step, variable-order solver based on the backward differentiation formula. 
The~\ac{LDSS} is translated into in equivalent model with fewer states, and a non-singular~$\mat{E}$ matrix using the~\texttt{isproper}~command in Matlab~\cite{MATLAB.2023}, which allows using the~\texttt{lsim} command to simulate the time response. 
A maximum step size of~\SI{1e-4}{\second}, and a relative tolerance of~\SI{1e-4}{} were chosen for both modeling approaches. 

The time-domain simulation of the three models are compared in the following for a small and a large disturbance induced by a small and a large load step, respectively. 
At the beginning of each simulation, the grid-forming and the grid-following model have the same active power reference to share the load equally.    

 \subsubsection{Small Load Step}
\begin{figure}
    \centering
    \includegraphics{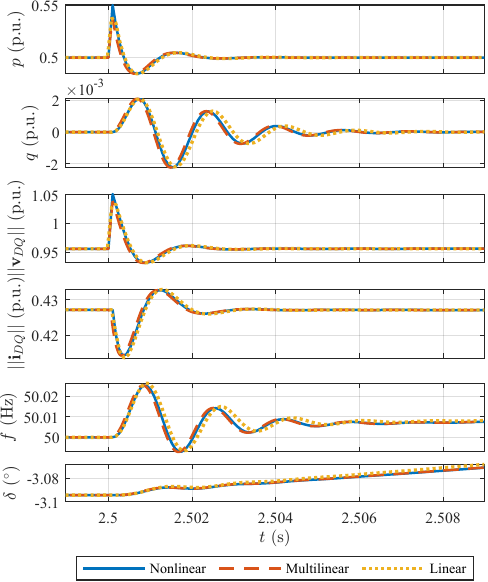}
     \caption{Response of the grid-following converter to a \SI{5}{\percent} load reduction at~\SI{2.5}{\second}}
     \label{fig:compSimRload5percent}
\end{figure}

A small load reduction of~\SI{5}{\percent} is triggered at~\SI{2.5}{\second}.
 The~\ac{iMTI} model is initialized and linearized~\SI{1}{\milli \second} before the disturbance.
The response of the grid-following converter is depicted in~Fig.~\ref{fig:compSimRload5percent}. 
The active and reactive power, $p$, and $q$, respectively, and the Euclidean norm of the voltage and the current, are plotted in per-unit, and the estimated frequency of the~\ac{PLL} in~\si{\hertz}, as well as the angle difference~$\delta$ between local and global frame, measured in Degrees, are given. 
The small load increase perturbs the active power injection of the~\ac{GFL} converter, which can be seen in all quantities. 
All three, the~\ac{NTI}~model, indicated by the~\ac{iMTI}, and the~\ac{LDSS}, indicated by the solid blue line, the red dashed, and the yellow dotted line, respectively, show a similar response. 
The~\ac{LDSS} deviates from the~\ac{NTI}~and the~\ac{iMTI} model, especially in the estimated frequency of the~\ac{PLL} during transients, with a small steady-state error after~\SI{2.507}{\second}. 
This is expected since the~\ac{LDSS} model only approximates the~\ac{iMTI} model, in particular when considering the multilinear model of the~\ac{PLL}.
The~\ac{iMTI} model is an exact representation of the~\ac{NTI}~model, as shown in Appendix~\ref{appx:PLLlieblack}, which can also be seen in the simulation results. 
The small deviations between the~\ac{NTI}~and the~\ac{iMTI} model are results of the different numerical solvers, where implicit solvers are generally more suitable when oscillations occur in the model~\cite{Cellier.2005}. 
One drawback of the~\texttt{ode15i} solver is that it can only simulate square models,~i.e., fully determined models with as many equations as unknowns. 
Therefore, the algebraic constraint~\eqref{eq:trigConstraint} is not considered in the numerical integration scheme, leading to a numerical drift~\cite{Ascher.1998}.
 
Overall, for the small disturbance the three models show the same response behavior, while the~\ac{LDSS} has a small steady-state deviation as a result of the approximation. 
\subsubsection{Large Load Step}
\begin{figure}
    \centering
    \includegraphics{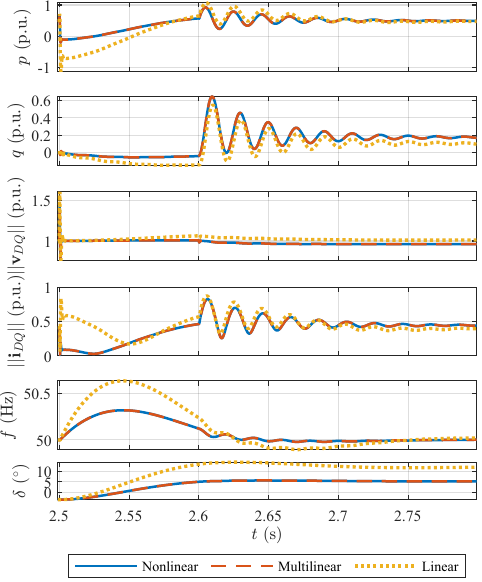}
    \caption{Response of the grid-forming converter to a \SI{50}{\percent} load reduction at \SI{2.5}{\second} and \SI{9}{\percent} voltage drop at \SI{2.6}{\second}}
    \label{fig:compRload50percent}
\end{figure}

The load reduction is now increased to~\SI{50}{\percent}. 
In addition, a voltage drop of~\SI{9}{\percent} occurs at~\SI{2.9}{\second}. 
The time response is depicted in~Fig.~\ref{fig:compRload50percent}. 
The larger disturbances results in the larger deviation from the equilibrium point, and consequently, the results of the~\ac{LDSS} model differ more than in the previous case. 
Besides the significant deviations during transients, the~\ac{LDSS} model also settles for a different steady state. 
In contrast, the~\ac{iMTI} model and the~\ac{NTI}~model show the same behavior as expected.

\subsubsection{Hopf Bifurcation}
\begin{figure}
    \includegraphics[width=0.95\columnwidth]{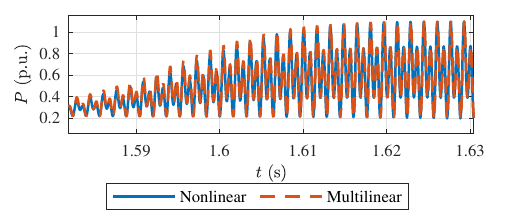}
    \caption{Example of a Hopf bifurcation seen in $P$ of the grid-forming converter}
    \label{fig:bifurcation}
\end{figure}

Power systems can experience bifurcations~\cite{Kundur.2004,Revel.2010}, which are nonlinear phenomena~\cite{Khalil.2015},~i.e., cannot be represented by~\ac{LSS}. 
The example network is driven on purpose to a point by increasing the active power setpoint to reach the power transfer limit, where a Hopf bifurcation starts to occur, shown in Fig.~\ref{fig:bifurcation}. 
The~\ac{iMTI} model matches the nonlinear model as expected.

\subsection{Small-Signal Analysis}
The generalized eigenvalues of the~\ac{LDSS} are compared with the eigenvalues of the~\ac{LSS} of the linearized~\ac{NTI}~model to assess the stability of the equilibrium point. 
The~\ac{LSS} is obtained using the~\texttt{linearize} command of the~\texttt{Control System Toolbox}. 
The~\texttt{linearize} command linearizes block by block using either an exact analytical linearization, or perturbs the right-hand side of the~\ac{ODE} within the block to linearize it numerically~\cite{MATLAB.2023}, where the default values are used.  
Both, the eigenvalues, and the generalized eigenvalues, are computed using the~\texttt{eig} command.

\begin{figure}
    \centering
    \begin{subfigure}[b]{\columnwidth}
    \includegraphics[width=\columnwidth]{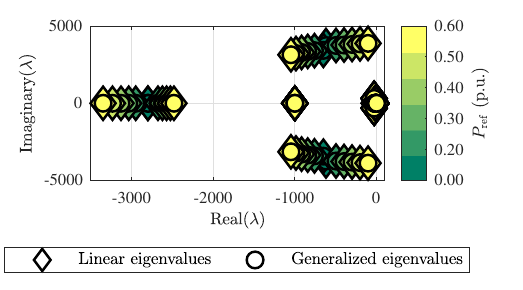}
    \caption{}
    \end{subfigure}
      \begin{subfigure}[b]{\columnwidth}
    \includegraphics[width=\columnwidth]{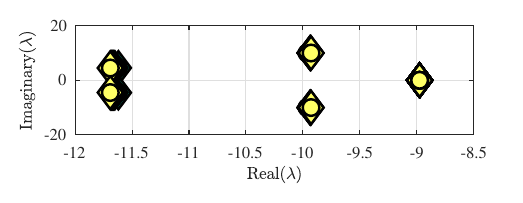}
    \caption{Zoom plot of (a)}
    \end{subfigure}    
    \caption{Comparison of the generalized eigenvalues of the linearized~\ac{iMTI} model with the eigenvalues of the linearized~\ac{NTI}~model}
    \label{fig:compEig}
\end{figure}

The models are linearized at~\SI{1}{\milli \second} before the load step, and the results are depicted in Fig.~\ref{fig:compEig}. 
The eigenvalues of the~\ac{LSS} are indicated by the diamond-shaped markers, and the finite generalized eigenvalues by the circles. 
The colormap indicates the active power set-point, where dark green indicates a low value, which increases with the brightness of the color to yellow. 
The~\ac{LDSS} has~\mbox{$\mathrm{rank}(\mat{E})=26$}~finite generalized eigenvalues, with some being zero.  
The non-zero generalized eigenvalues match the eigenvalues of the~\ac{LSS}, which is expected as mentioned in the previous section. 
Small deviations occur due to the numerical methods of computing the eigenvalues, as well as the numerical approximation of the Jacobian for parts of the~\ac{NTI}~model in contrast to the exact linearization of the~\ac{iMTI} model. 

The small-signal analysis of the~\ac{iMTI} model comes to the same conclusion as the linearized~\ac{NTI}~model, that the equilibrium points are stable, since there are no eigenvalues with positive real part. 

\subsection{Computational Assessment}
The computation time of the proposed linearization method for~\ac{iMTI} models is compared with the linearization using the~\texttt{Control System Toolbox} from Simulink. 
The code in Matlab 2023b~\cite{MATLAB.2023} was executed on a x64-based laptop with Microsoft Windows 11, an~\mbox{Intel® Core™ i7-1370P} processor, and~\SI{32}{\giga \byte}~RAM. 
Table~\ref{tab:computationTime} compares the computation time for linearization~(lin) and eigenvalue calculation~(eig), where the model properties indicate the number of states~$n$, the number of inputs~$m$, the number of algebraic and output variables~$Q$, and the number of factors of the~\ac{iMTI} model~$R$.  
The linearization using the~\texttt{MTI Toolbox} is significantly faster than the approach using~\texttt{Control System Toolbox}, since only the operating point needs to be inserted.
The computation of the generalized eigenvalues requires more time than for the computation of eigenvalues in the studied cases. 
The increase of the size of the~\ac{iMTI} model lead to an increase in computation time as expected. 
Therefore, for the studied cases, the linearization of the~\ac{iMTI} model was more efficient than the~\texttt{Control System Toolbox}. 

\begin{table}
    \caption{Comparison of the computation time of the linearization (lin) and the eigenvalue computation (eig) for different configuration: a)~\ac{GFM} b)~\ac{GFL} c) \ac{GFM}+\ac{GFL} }
    \label{tab:computationTime}
     \renewcommand{\arraystretch}{1.5} 
    \begin{tabularx}{\columnwidth}{X X X p{0.4cm} X c c c c }
        \hline
        \multicolumn{5}{c}{\shortstack{\textbf{Model properties} \\ \phantom{Toolbox}}} & \multicolumn{2}{c}{ \shortstack{\textbf{\texttt{Control System}} \\ \textbf{\texttt{Toolbox}}}}& \multicolumn{2}{c}{ \shortstack{\textbf{\texttt{MTI Toolbox}} \\ \phantom{\texttt{Toolbox}}}}\\
      &$n$ & $m$& $\!p\!+\!q\!$ &$R$ &  lin~(\si{\milli\second})  & eig~(\si{\milli\second}) & lin~(\si{\milli\second}) & eig~(\si{\milli\second})  \\
        \hline
       a)& $16$ & $6$&  $17$&$87$ & $309.7$   & $0.123$  &  $0.0133$   &$0.1420$ \\
       b)& $15$&$6$ & $17$ & $95$ &$450.34$  & $0.1091$  &   $0.0144$   &$0.1221$ \\
       c)&$26$&$8$ &$32$ & $165$  & $471.02$ & $0.257$ & $0.0474$ &$0.4776$\\
        \hline
    \end{tabularx}
\end{table}

%% file: content/sec_conclusions.tex
\section{Conclusion}\label{sec:con}
This paper proposes an efficient approach to perform small-signal power system stability analysis using~\acl{iMTI} models, with a focus on converter-dominated networks. 
It explains how~\ac{iMTI} models are represented by~\ac{CPN1}~decomposed tensors, and how some nonlinear functions can be written as an~\ac{iMTI} model. 
The~\ac{iMTI} model is an exact representation of the nonlinear model, as the paper shows that the change of variables used to represent trigonometric functions is a Lie-Bläcklund transformation,~i.e., the models are equivalent. 
This equivalence could be seen in both, the time-domain simulation results, and the small-signal stability analysis of a 3 bus-network, where the proposed approach was compared with the nonlinear modeling approach. 
To enable the small-signal analysis of~\ac{iMTI} models, this paper introduced the efficient linearization of~\ac{iMTI} models, where the~\ac{iMTI} is approximated locally by an~\ac{LDSS} model. 
From the~\ac{LDSS} model, the generalized eigenvalues were computed to assess the stability of the equilibrium point. 
A comparison of the computational time also showed that the factorized format of the~\ac{CPN1} tensor decomposition of the~\ac{iMTI} model enables an efficient linearization compared to the~\texttt{linearize} command provided by Matlab, with similar degree of accuracy. 

Future research will extend the assessment of the proposed~\ac{iMTI} modeling of power systems to larger power networks. 
The representation as decomposed tensors makes the multilinear modeling class a promising candidate for the modeling and analysis of large-scale low-inertia, converter-dominated networks.
Finding lower rank approximations for the multilinear power system models is a key challenge for this.
Additionally, the power system models can be extended to include more details, such as saturations, which can easily be represented as~\ac{iMTI} models~\cite{Lichtenberg.2022}. 
Furthermore, other approaches for the stability analysis of~\ac{iMTI} models of power systems beyond small disturbances will be explored in the future.

%% file: content/backmatter.tex
\section*{Acknowledgments}
The authors would like to thank Irina Subotic, who pointed at the work of Matteo Tacchi.

 The authors’ views do not necessarily reflect those of the European Union.

\input{content/appendix.tex}

\section{References Section}

\bibliographystyle{IEEEtran}

\bibliography{bib/main.bib}

\section{Biography Section}

\begin{IEEEbiography}[{\includegraphics[width=1in,height=1.25in,clip,keepaspectratio]{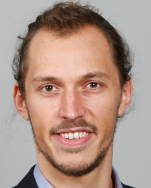}}]{Christoph Kaufmann}
 received his B.Sc. in Industrial Engineering and Management from the University of Hamburg in 2016, and his M.Eng. in Renewable Energy Systems at the HAW Hamburg in 2019. He is working as a Research Associate at the Fraunhofer Institute for Wind Energy Systems IWES and currently pursuing a PhD at the Technical University of Catalonia and the Hamburg University of Applied Sciences on stability analysis of power systems using on multilinear models.
\end{IEEEbiography}

\begin{IEEEbiography}[{\includegraphics[width=1in,height=1.25in,clip,keepaspectratio]{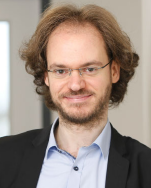}}]{Georg Pangalos}
received the Dipl.-Ing. degree in Electrical Engineering and the Ph.D. degree in Control Engineering from Hamburg University of Technology in 2011 and 2015, respectively. Since 2015, he has been with Fraunhofer-Gesellschaft zur Förderung der angewandten Forschung e.V., Hamburg, Germany. His research interests include renewable energy integration and energy systems transition.
\end{IEEEbiography}

\begin{IEEEbiography}[{\includegraphics[width=1in,height=1.25in,clip,keepaspectratio]{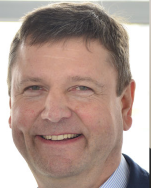}}]{Gerwald Lichtenberg}
studied Physics at the Universities of Heidelberg and Hamburg, received his Ph.D. in Control Engineering in 1998 and habilitation in Systems Theory in 2012, both from Hamburg University of Technology. Since 2013, he has been Professor of Physics and Control at Hamburg University of Applied Sciences and Scientific Director of the Fraunhofer Application Center for Integration of Local Energy Systems. His research focuses on multilinear modeling and tensor decomposition applications for control problems.
\end{IEEEbiography}

\begin{IEEEbiography}[{\includegraphics[width=1in,height=1.25in,clip,keepaspectratio]{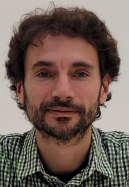}}]{Oriol Gomis-Bellmunt} (Fellow, IEEE)
received the degree in Industrial Engineering from the Technical University of Catalonia (UPC) in 2001 and the Ph.D. in Electrical Engineering from UPC in 2007. He joined Engitrol S.L. in 1999 as a Project Engineer in automation and control. Since 2004, he has been a Professor in the Electrical Engineering Department at UPC and participates in the CITCEA-UPC Research Group. He has been an ICREA Academia researcher since 2020 and co-founded the start-up eRoots Analytics in 2022, focusing on modern power systems analysis. His research interests include power electronics, power systems, and renewable energy integration.
\end{IEEEbiography}

\vfill

%% file: content/appendix.tex
\appendix
 {\appendices

\section{Equivalence of trigonometric nonlinearities} \label{apx:equiDyn}
\subsection{Lie-Bläcklund Transformation} \label{sec:LBtrf}
Lie-Bläcklund transformations ensure the equivalence of the dynamics of the finite-dimensional model
\begin{IEEEeqnarray}{rCl}
\dot{\bm{\xi}}&=&\vec{f}(\bm{\xi}) \,,
\end{IEEEeqnarray}
with vector field~$\vec{f}$
defined on the smooth manifold~$\mathcal{M}$. 
A Lie-Bläcklund transformation is defined as follows~\cite{Fliess.1999,Tacchi.2018}.

\begin{definition}{\cite{Fliess.1999,Tacchi.2018}}
Let~$\left(\mathcal{M}, \vec{f}\right)$, and~$\left(\mathcal{N}, \vec{h}\right)$, be two models, $\mat{\TrfVars}:\,\mathcal{M}\stackrel{C^\infty}{\to} \mathcal{N}$, $p \in \mathcal{M}$, and $\vec{q}:=\TrfVars(\vec{p})\in\mathcal{N}$. Then,
\begin{itemize}
    \item if $\bm{\xi}$ is a trajectory of~$(\mathcal{M}, \vec{f})$ in a neighborhood of~$\vec{p}$, then $\bm{\zeta} := \mat{\TrfVars} \circ  \bm{\xi}$ stays in a neighbourhood of q, and 
    \begin{IEEEeqnarray}{rCl}
        \dot{\bm{\zeta}}(t)&=&\nabla \mat{\TrfVars}\left(\bm{\xi}(t)\right) \cdot \vec{f}\left(\bm{\xi}(t)\right) \,,
    \end{IEEEeqnarray}
    which holds even in infinite dimensions: everything depends only on a finite number of coordinates. 
    \item $\mat{\TrfVars}$ is an endogenous transformation if and only if, for any~$\bm{\xi}$ in a neighbourhood of~$p$
     \begin{IEEEeqnarray}{rCl}
        \nabla\mat{\TrfVars}\left(\bm{\xi}\right) \cdot \mat{f}\left(\bm{\xi}\right)&=& \mat{h}\left( \mat{\TrfVars}(\bm{\xi})\right)\,, \label{eq:endTrf}
    \end{IEEEeqnarray}
    it is said that~$\mat{f}$ and~$\mat{h}$ are~$\TrfVars$-related at $(\vec{p}, \vec{q})$; then~\mbox{$\dot{\bm{\zeta}}=\mat{h}(\bm{\zeta})$} and~$\bm{\TrfVars}$ has a smooth inverse~$\bm{\Psi}$. Consequently,~$\mat{h}$ and~$\mat{f}$ are~$\Psi$-related. 
    \item $\mat{\TrfVars}$ is a Lie-Bläcklund isomorphism if and only if it holds locally that
    \begin{IEEEeqnarray}{rCl}
    \mat{T}\mat{\TrfVars}\left(\text{span}(\mat{f})\right)=\text{span}\left( \mat{h}\right)\,,
    \end{IEEEeqnarray}
    and $\vec{\TrfVars}$ has smooth inverse~$\bm{\Psi}$ such that $\mat{T}\mat{\TrfVars}\left(\text{span}(\mat{h})\right)=\spn(\mat{f})$. That is, for $\bm{\xi}$ in a neighborhood of $\mat{p}$, and $\bm{\zeta}$ in a neighborhood of~$\mat{q}$, it should hold that
    \begin{IEEEeqnarray}{rCl}
        \nabla  \mat{\TrfVars}(\bm{\xi}) \cdot (\spn(\mat{f}(\bm{\xi})))&=& \spn(\mat{h}( \vec{\TrfVars}(\bm{\xi})))\,,\\
         \nabla  \mat{\Psi}(\bm{\zeta}) \cdot (\spn(\mat{h}(\bm{\zeta})))&=& \spn(\mat{f}( \vec{\Psi}(\bm{\zeta})))\,.
    \end{IEEEeqnarray}
\end{itemize}

\end{definition}

\subsection{Proof of Equivalence of the PLL dynamics} 
\label{appx:PLLlieblack}
The following shows that the transformation~\eqref{eq:PLLtmap} used to express the~\ac{PLL} as a multilinear model preserves the dynamics, meaning that the stability assessment is valid for the nonlinear case despite the change of state dimensions. 

It needs to be checked if it is an endogenous transformation between the model~$(\mathbb{R}^2,\vec{f})$ and~$(\mathcal{M},\vec{h})$.
This is done by checking if the transformation~$\TrfVars$ has a smooth inverse~$\Psi$.
Furthermore, from~\eqref{eq:endTrf} we need to check~\mbox{$\vec{h}(\TrfVars(\vec{x}),\vec{u})=\nabla \TrfVars(\vec{x}) \vec{f}(\vec{x},\vec{u})$}. 
The procedure from~\cite{Tacchi.intern.2017} is followed. 

By checking if $g\circ \bm{\TrfVars}=0$, we can check if the transformation $\bm{\TrfVars}$ is well-defined. Calculating  
\begin{IEEEeqnarray*}{rCl}
g\left(\TrfVars(\vec{x})\right)&=&\sin^2{(x_1)}+\cos^2{(x_1)}-1\,,\\
&=&1-1=0\,,
\end{IEEEeqnarray*}
shows that~$\TrfVars$ is well-defined. 

Then we check if $\bm{\Psi}$ is a smooth inverse to $\bm{\TrfVars}$. The inverse transformation is defined in terms of complex variables as 
\begin{IEEEeqnarray}{rCl}
    \Psi(\vec{z})&=&\left(\arg(z_1 + i z_2), z_3 \right)\,,
\end{IEEEeqnarray}
where $i$ is the imaginary unit. Each argument is checked if it is well-defined and smooth in the following. 

The angle~$\arg(z_1 + i z_2)$ is well-defined and smooth, if and only if~\mbox{$z_1 + i z_2\neq0 $}. 
The constraints~\eqref{eq:PLL_dq_constraints} with~\mbox{$g(\vec{z})=z_1^2+z_2^2 -1=0$} forces, when~$z_1=0$, then~\mbox{$z_2=\pm1$}, and vice versa. 
The equivalence constraints~\eqref{eq:PLL_dq_constraints} are smooth and well-defined. 
Hence,~$\Psi$ is well-defined.
Furthermore,~$\Psi$ is smooth as it is~$C^k$,~i.e.,~$k$-times differentiable.
 Replacing~$\vec{z}$ with the transformation~$\TrfVars(\vec{x})$ leads to the original states
\begin{IEEEeqnarray*}{rCl}
    \Psi\left(\TrfVars(\mathbf{x})\right)&=&\left(\arg\left(\cos(x_1) + i \sin(x_1)\right), x_2,u_1,u_2\right)\,,\\
    &=&\left(x_1, x_2,u_1,u_2 \right)\,.
\end{IEEEeqnarray*}

Lastly, it needs to be shown that 
\begin{IEEEeqnarray*}{rCl}
    \vec{h}\left( \TrfVars(\mathbf{x})\right)=\nabla  \TrfVars(\vec{x}) \cdot \vec{f}(\vec{x},\vec{u})\,,
\end{IEEEeqnarray*}
where 
\begin{IEEEeqnarray*}{rCl}  
  \vec{h}\left( \TrfVars(\mathbf{x})\right)\!\!=\!\!\begin{pmatrix}-\left(x_2-k_\mathrm{p}\left(u_1 \sin{(x_1)}+u_2\cos{(x_1)}\right) \right)\sin{(x_1)}\\
\left(x_2-k_\mathrm{p}\left(u_1\sin{(x_1)}+u_2 \cos{(x_1)}\right) \right) \cos{(x_1)} \\
-k_\mathrm{i}\left(u_1\sin{(x_1)}+u_2\cos{(x_1)}\right)  \end{pmatrix}\!\!,
\end{IEEEeqnarray*}
and 
\begin{IEEEeqnarray*}{rCl} 
    \renewcommand{\arraystretch}{1.1}
 \nabla  \bm{\TrfVars}(\vec{x})=\begin{pmatrix}
 - \sin{(x_1)} & 0 & 0 &0 \\
  \cos{(x_1)}& 0 & 0 & 0\\
 0 & 1 & 0 & 0\\
  0 & 0 & 1 & 0\\
   0 & 0 & 0 & 1\\
\end{pmatrix}\,,
\end{IEEEeqnarray*}
and 
\begin{IEEEeqnarray*}{rCl}
    \renewcommand{\arraystretch}{1.1}
    \vec{f}(\vec{x},\vec{u})=\begin{pmatrix}x_2-k_\mathrm{p}\left(u_1\sin{(x_1)}+u_2\cos{(x_1)}\right)   \\
-k_\mathrm{i}\left(u_1\sin{(x_1)}+u_2\cos{(x_1)}\right)\\
0\\
0\end{pmatrix}\,,
\end{IEEEeqnarray*}
which is true. 

Thus, the transformation~\eqref{eq:PLLtmap} is an endogenous transformation of the Lie-Bläcklund transformation. 

} 